\documentclass[11pt,a4paper]{article}
\pdfoutput=1

\usepackage{jheppub}
\usepackage[T1]{fontenc}
\usepackage{extarrows}

\title{Linearized fluid/gravity correspondence: from shear viscosity to all order  hydrodynamics}
\author{Yanyan Bu}
\author{and Michael Lublinsky}
\affiliation{Department of Physics, Ben-Gurion University of the Negev, \\
Beer-Sheva 84105, Israel}

\emailAdd{yybu@post.bgu.ac.il}
\emailAdd{lublinm@bgu.ac.il}

\abstract{In ref.~\cite{1406.7222}, we reported a construction of all order linearized fluid dynamics with strongly coupled $\mathcal{N}=4$ super-Yang-Mills theory as  underlying microscopic description. The linearized fluid/gravity correspondence makes it possible to resum all order derivative terms in the fluid stress tensor. Dissipative effects are fully encoded by the shear term and a new one, emerging starting from third order in hydrodynamic derivative expansion. In this work, we provide all computational details omitted in~\cite{1406.7222} and present additional results. We derive closed-form linear holographic RG flow-type equations for momenta-dependent transport coefficient functions. Generalized Navier-Stokes equations are shown to emerge from the constraint components of the bulk Einstein equations. We perturbatively solve the RG equations for the viscosity functions, up to third order in derivative expansion, and up to this order compute spectrum of small fluctuations. Finally, we solve the RG equations numerically,
thus accounting for all order derivative terms in the boundary stress tensor.}


\arxivnumber{1409.3095}
\begin{document}
\maketitle

\flushbottom

\section{Introduction}\label{section1}
In heavy ion collisions at RHIC and LHC, a novel state of QCD matter, quark-gluon plasma, is created. The quark-gluon plasma produced at RHIC was discovered to behave like a nearly perfect fluid reflecting strongly coupled regime of QCD~\cite{nucl-th/0405013,hep-ph/0405066}, with relativistic hydrodynamics found to provide an accurate description of the plasma fireball expansion. The hydrodynamic evolution of the quark-gluon plasma is characterized by a set of transport coefficients, which have to be computed from the microscopic QCD. However, strongly coupled nature of this system prevents from a first principe analytic calculation of these coefficients. While lattice methods are quite reliable in studying QCD thermodynamics~\cite{hep-lat/0611014,1007.2580,1309.5258}, they usually fail in extracting transport coefficients due to limited applicability to real-time dynamics. Therefore, various microscopic models are indispensable to understand transport properties of this QCD matter.

Fluid dynamics~\cite{fluid1,fluid2} is an effective description of most interacting quantum field theories at long wavelengths. The description is of statistical nature: it captures collective dynamics of a large number of microscopic degrees of freedom. The collective variables suitable for such a description are local densities of conserved charges, local fluid velocity and temperature. Hydrodynamic equations are local conservation laws for corresponding currents, which are specified via constitutive relations. As a low energy effective field theory, fluid dynamics describes near-thermal equilibrium systems and is naturally defined in terms of derivative expansion of the local fluid mechanical variables. Up to a finite number of transport coefficients, the derivative expansion at any given order is completely fixed by thermodynamic and symmetry considerations. Transport coefficients, such as viscosities and conductivities, must be determined from either experimental measurements or theoretical computations in the underlying microscopic theory.

The stress tensor $T_{\mu\nu}$ of a relativistic fluid is usually presented as a sum of two terms
\begin{equation}\label{full stress tensor}
T_{\mu\nu}=T_{\mu\nu}^{\text{Ideal}}+T_{\mu\nu}^{\text{Diss}},
\end{equation}
where $T_{\mu\nu}^{\text{Ideal}}$ corresponds to ideal fluid dynamics and for conformal fluids has the form\footnote{The central charge of boundary CFT was factored out by a suitable normalization of the Boltzmann constant $k_{\textrm{B}}$.}
\begin{equation}\label{ideal stress tensor}
T_{\mu\nu}^{\text{Ideal}}=\frac{1}{{\bf b}^4}(\eta_{\mu\nu}+4u_{\mu}u_{\nu}),
\end{equation}
where $\eta_{\mu\nu}$ is the four dimensional Minkowski metric, the fluid's four velocity is $u_\mu$ and, in our notations, the temperature is
\begin{equation}
T=\frac{1}{\pi \bf b}.
\end{equation}
$T_{\mu\nu}^{\text{Diss}}$ accounts for dissipative effects and is chosen to be nonzero only for spatial components. Up to the first velocity gradient it is given by the Navier-Stokes term\footnote{The bulk viscosity vanishes for the conformal fluids to be discussed below.}
\begin{equation}
T_{ij}^{\text{Diss}}=-\eta_0\,\sigma_{ij},
\end{equation}
where $\eta_0$ is the shear viscosity coefficient and the tensor $\sigma_{ij}$ has the
form\footnote{The linearization~(\ref{linearization}) was assumed in writing down the shear tensor $\sigma_{ij}$.}
\begin{equation}\label{sigma}
\sigma_{ij}=\frac{1}{2}\left(\partial_i\beta_j+\partial_j\beta_i-\frac{2}{3}\delta_{ij} \partial\beta\right),~~~~~~~~~u_i=\frac{\beta_{i}}{\sqrt{1-\beta^2}}.
\end{equation}

Theoretical foundations of relativistic dissipative fluid dynamics are not yet fully established. The relativistic Navier-Stokes equations are a-causal and unstable~\cite{instable1,instable2,instable3,instable4}: the irreversible currents are linearly proportional to the thermodynamic forces, which have instantaneous influence on the currents, obviously violating causality. These problems can be solved by introducing retardation into the definitions of the irreversible currents~\cite{IS1,IS2}, leading to equations of motion for these currents, which thus become independent dynamical variables. Theories of this type are known as causal relativistic dissipative fluid dynamics. Causality usually also implies stability~\cite{0907.3906}. To obtain a causal formulation, one needs to include higher order terms in the gradient expansion of the currents, in which case additional transport coefficients arise. Truncation at any fixed order would presumably lead to violation of causality that can be fully restored only at infinite order, which we refer to as an all-order gradient resummation. Non-trivial physical consequences imposed by causality of relativistic fluids were investigated
in~\cite{instable1,instable2,instable3,instable4,0807.3120,0907.3906,1102.4780} and lead to certain constraints on possible values of higher order transport coefficients.

AdS/CFT correspondence~\cite{hep-th/9711200,hep-th/9802109,hep-th/9802150,hep-th/9905111} emerged over a decade ago as a standard tool and a model playground for addressing strongly coupled dynamics of gauge theories. The AdS/CFT correspondence relates holographically a large $N$ strongly interacting quantum field theory with dynamics of classical gravity in (asymptotically) AdS spacetime. Fluid/gravity correspondence~\cite{hep-th/0104066,hep-th/0205052,hep-th/0405231,hep-th/0512162} is a long wavelength limit of the AdS/CFT correspondence: it gives a map between black holes in asymptotically AdS spacetime and fluid dynamics of a strongly coupled boundary field theory. The most celebrated prediction of the fluid/gravity correspondence is a ratio of the shear viscosity $\eta_0$ to the entropy
density $s$~\cite{hep-th/0104066,hep-th/0205052,hep-th/0405231},
\begin{equation}\label{ratio}
\frac{\eta_0}{s}=\frac{1}{4\pi}.
\end{equation}
The ratio~(\ref{ratio}) is \emph{universal} for a large class of strongly coupled gauge theory plasmas for which  holographic duals are governed by Einstein gravities in asymptotically AdS spacetimes. Remarkably, this small ratio is quite close to the values extracted for QCD plasma from the RHIC experiments. Universality of this ratio was further proved in~\cite{hep-th/0311175,0808.1837} and in~\cite{0809.3808} using black hole membrane paradigm. It was later found that university of the ratio~(\ref{ratio}) gets violated by either modifying Einstein gravity~\cite{0712.0743,0712.0805,0806.2156,
0812.2521,0811.1665,0901.1421,0903.2527,0903.2834,0903.3244,0908.1473} or breaking isotropic invariance among spatial directions~\cite{1011.5912,1109.4592,1110.6825,1406.6019}. We refer the reader to~\cite{0704.0240} for a comprehensive review on early literature.

A specific realization of the fluid/gravity correspondence, and the one we will closely follow below, was established in~\cite{0712.2456}: it provides a systematic framework to construct a universal nonlinear fluid dynamics, order by order in the boundary derivative expansion. The stress tensor for the dual conformal fluid was explicitly constructed in~\cite{0712.2456} up to second order in derivative expansion, which is in perfect agreement with a general form of second order conformal hydrodynamics as analyzed in~\cite{0712.2451}. Computations of~\cite{0712.2456} were later generalized to conformal fluids in flat~\cite{0806.4602} and also weakly curved~\cite{0809.4272} background manifolds of various dimensions. Forced fluids in a weakly curved manifold were examined in~\cite{0806.0006} by studying long wavelength solutions of Einstein-dilaton gravity with negative cosmological constant. Further developments can be found in reviews~\cite{0905.4352,1107.5780} and references therein.

Refs.~\cite{0704.1647,0905.4069} (see also~\cite{1103.3452,1203.0755,1302.0697}) initiated study of generalized relativistic hydrodynamics by considering all orders in derivatives of local fluid mechanical variables in the stress tensor.

All order or resummed hydrodynamics is found to accommodate certain contributions, which are not present in a strict low momentum approximation. To avoid any confusion we thus clarify the terminology: by hydrodynamics we mean an effective theory given by a constitutive relation for the stress tensor in terms of temperature, fluid velocity and their gradients only. Navier-Stokes or any ``unresummed'' hydrodynamics involves only a finite number of gradients, while all order or ``resummed'' hydrodynamics means infinite number of gradients, but no other degrees of freedom.

The higher order derivative expansion generally includes two types of terms: nonlinear in the fluid velocity, like $(\nabla u)^2$, and linear ones like $\nabla\cdots\nabla u$. The nonlinearities are significant when amplitudes of local fluid mechanical variables are large. However, even for fluid perturbations with small amplitudes, one can get large  contributions from the linear terms when the momenta associated with the fluid perturbations are large. Given that these two types of terms are controlled by different parameters, it is possible to separate these contributions and to have the linear terms under theoretical control. We will focus on those linear terms in the rest of this paper.

To collect all order linear terms in a self-consistent manner, the shear viscosity $\eta_0$ was generalized into momenta-dependent function $\eta(\omega,q^2)$. This viscosity function is expressed in momentum space which follows from the replacement $\partial_{\mu}\to \left(-i\omega,i\vec{q}\right)$ in the linear gradient expansion of $T_{\mu\nu}$. By postulating a constitutive relation for $T_{\mu\nu}$ in terms of the shear viscosity function $\eta(\omega,q^2)$, the authors of~\cite{0905.4069} attempted to extract $\eta(\omega,q^2)$ from thermal correlators of the stress tensor computed on the gravity side~\cite{hep-th/0205051,hep-th/0602059}. While certain progress was achieved in~\cite{0905.4069}, its prime goal of complete determination of the viscosity function  was not reached. It was realized that even in the case of linearized hydrodynamics, knowledge of retarded correlators contains insufficient information about all transport properties of the system. In ref.~\cite{1406.7222}, we succeeded to completely solve this problem and below we provide all the details related to our computations.

In ref.~\cite{1406.7222}, we reported progress, achieved via linearizing fluid/gravity correspondence, in consistently generalizing relativistic hydrodynamics to all orders. Upon linearization, perturbative computations of~\cite{0712.2456} can be straightforwardly extended to arbitrary order in the boundary derivative expansion. Our procedure is, however, slightly different from that of~\cite{0712.2456}. Particularly, the dissipative part in the stress tensor is collected in a unified way, rather than being determined order by order in derivative expansion. The Einstein equations in the bulk are split into two sets: dynamical equations and constraints. It turns out that in order to derive transport coefficient functions, it is sufficient to solve dynamical components of the bulk Einstein equations only. By solving only those we construct an ``off-shell'' fluid stress tensor. The remaining constraint components of the Einstein equations are equivalent to conservation laws of thus constructed fluid stress tensor and lead to generalized Navier-Stokes equations. It is worth emphasizing that the bulk dynamics is not absorptive, rather the bulk acts as a non-linear dispersive medium. Dissipative effects emerge via absorptive boundary conditions at the black hole horizon.

We find that the dissipative part of the stress tensor has the following form
\begin{equation}\label{stress tensor nonperturb}
T_{ij}^{\text{Diss}}=-\left[\eta(\omega,q^2)\sigma_{ij} + \zeta(\omega,q^2)\pi_{ij}\right],
\end{equation}
where $\pi_{ij}$ is a third order tensor structure
\begin{equation}\label{tensor structures}
\pi_{ij}=\partial_i\partial_j\partial\beta-\frac{1}{3}\delta_{ij}\partial^2\partial\beta
\end{equation}
and $\zeta(\omega,q^2)$ is a new viscosity function, which in~\cite{0905.4069}, apparently incorrectly, was argued to be zero. We here express the viscosity functions in momentum space but with tensors $\sigma_{ij}$ and $\pi_{ij}$ formulated as explicit derivatives of the fluid velocity. Later, the notations $\eta(\partial_v,\partial^2)$ and $\zeta(\partial_v,\partial^2)$ will also be used to denote the viscosity functions. All these expressions are interchangeable under the above rule of replacement. We will be working using dimensionless units for all the momenta, choosing units such that the temperature is normalized to $\pi T=1$.  So all the physical momenta should be understood as in units of $\pi\,T$: $\omega \pi T$ and $q_i \pi T $.

In the hydrodynamic limit $\omega,q_i\ll 1$, $\eta(\omega,q^2)$ and $\zeta(\omega,q^2)$ are expandable in power series and a perturbative
analysis to be presented below reveals a few first terms in the expansions,
\begin{equation}\label{viscosity funs}
\begin{split}
\eta(\omega,q^2)&=2+(2-\ln{2})i\omega-\frac{1}{4}q^2-\frac{1}{24}\left[6\pi-\pi^2+ 12\left(2-3\ln{2}+\ln^2{2}\right)\right]\omega^2+\cdots,\\
\zeta(\omega,q^2)&=\frac{1}{12}\left(5-\pi-2\ln{2}\right)+\cdots,
\end{split}
\end{equation}
where in our units, the first term in $\eta$ corresponds to the ratio~(\ref{ratio}), while the second term  accounts for the relaxation time~\cite{hep-th/0703243,0712.2451,0712.2456,0712.2916,0811.1794}. The remaining two terms in equation~(\ref{viscosity funs}) are new third order transport coefficients.
In section~\ref{section4}, $\eta(\omega,q^2)$ and $\zeta(\omega,q^2)$ are computed numerically to all orders, and we will  also present
expansions of these viscosity functions up to fifth order.

The remaining part of this paper is organized as follows. In section~\ref{section2}, we outline the derivation of the fluid dynamics from the bulk gravity. The main results are formulated as closed linear holographic RG flow-type equations for the generalized transport coefficient functions and generalized Navier-Stokes equations. We then compute  dispersion relations for sound and shear modes. In section~\ref{section3}, to make comparison with previous studies in the literature, we perturbatively solve these holographic RG flow-type equations and obtain the fluid stress tensor up to third order in the derivative expansion. In section~\ref{section4}, we numerically solve the RG flow-type equations and obtain generalized transport coefficient functions, extending the perturbative analysis of section~\ref{section3} to very large momenta. We also extract the viscosity functions via an approximate matching scheme, and find perfect agreement with the numerical results. Summary and discussion can be found in section~\ref{section5}.
\section{Fluid dynamics from the bulk gravity}\label{section2}
\subsection{Linearized fluid/gravity correspondence}\label{subsection21}
The fluid/gravity correspondence makes it possible to construct the fluid stress tensor and prove its conservation law (Navier-Stokes equations) by solving the bulk Einstein equations in asymptotically AdS spacetime, in the long wavelength limit. We start by considering a universal sector of the AdS/CFT correspondence: classical Einstein gravity with a negative cosmological constant in five dimensional spacetime,
\begin{equation}\label{action}
S=\frac{1}{16\pi G_{N}}\int d^5x \sqrt{-g}\left(R+12\right),
\end{equation}
where the AdS radius is set to one for convenience. The Einstein equations which follow from the action~(\ref{action}) are
\begin{equation}\label{einstein eq}
E_{MN}\equiv R_{MN}-\frac{1}{2}g_{MN}R-6g_{MN}=0.
\end{equation}
We use upper case Latin indices $\{M,N,\cdots\}$ and lower case Greek indices $\{\mu,\nu,\cdots\}$ to denote bulk and boundary directions, respectively. Lower case Latin indices $\{i,j,\cdots\}$ will be used to specify spatial directions along the boundary.

Besides pure $AdS_5$ spacetime which is dual to the vacuum state of the boundary CFT, the action~(\ref{action}) also admits a 4-parameter family of Black Hole solutions,
\begin{equation}\label{boosted bh}
ds^2=-2u_{\mu}dx^{\mu}dr-r^2f\left({\bf b} r\right) u_{\mu}u_{\nu}dx^{\mu}dx^{\nu}+r^2 \mathcal{P}_{\mu\nu} dx^{\mu}dx^{\nu},
\end{equation}
with
\begin{equation}\label{velocity}
u_v=-\frac{1}{\sqrt{1-\beta^2}},~~~~u_i=\frac{\beta_{i}}{\sqrt{1-\beta^2}},~~~~ \beta^2=\sum_{i=1}^{3}\beta_i\beta_i,
\end{equation}
and $f(r)=1-1/r^4$. Hawking temperature of this black hole is
\begin{equation}
T=\frac{1}{\pi \bf b}
\end{equation}
which is identified as the temperature of the dual  CFT. The operator $\mathcal{P}_{\mu\nu}=\eta_{\mu\nu}+u_{\mu}u_{\nu}$ acts as a projector onto spatial directions. Notice that so far the parameters $\beta_i$ and $\bf b$ are held constant, so that the line element~(\ref{boosted bh}) does form a class of solutions to the Einstein equations~(\ref{einstein eq}). As pointed out in~\cite{0712.2456}, the line element~(\ref{boosted bh}) defines a metric of a uniform black brane written in the ingoing Eddington-Finkelstein coordinate, moving at velocity $\beta_i$ along spatial direction $x_i$.

Discussing fluid dynamics we closely follow~\cite{0712.2456} and promote the constant parameters $\beta_i$ and $\bf b$ into arbitrarily slowly varying functions of boundary coordinates $x^{\alpha}$,
\begin{equation}\label{metric1}
ds^2=-2u_{\mu}(x^{\alpha})dx^{\mu}dr-r^2f\left({\bf b}(x^{\alpha}) r\right) u_{\mu}(x^{\alpha})u_{\nu}(x^{\alpha})dx^{\mu}dx^{\nu}+r^2 \mathcal{P}_{\mu\nu}(x^{\alpha}) dx^{\mu}dx^{\nu},
\end{equation}
In general, (\ref{metric1}) no longer solves the Einstein equations~(\ref{einstein eq}). The method developed in~\cite{0712.2456} is to add suitable corrections to~(\ref{metric1}), so that~(\ref{einstein eq}) is satisfied by the new line element. The corrected metric is not easily found for a general configuration. The authors of~\cite{0712.2456} introduced a systematic way to construct the corrected metric, expanding in the long wavelength limit. A scale associated with this expansion should be much larger than a characteristic scale of the system, such as inverse of the temperature $1/T$. That is the velocity and temperature fields are assumed to vary slowly on this scale, admitting a gradient expansion around some arbitrarily chosen point, such as spacetime origin $x^{\alpha}=0$. The Einstein equations~(\ref{einstein eq}) for the metric are then solved order by order in the boundary derivative expansion. As has been pointed out in the Introduction, the dual metric was constructed up to second order in the velocity gradient, including nonlinear terms quadratic in the velocity gradient.

Our main goal is to perform a summation over all higher order derivative terms in the boundary stress tensor, but as stressed employing linear approximation. Rather than resorting to order by order derivative expansion of~\cite{0712.2456}, we instead linearize the problem in perturbations of the fluid mechanical variables $u_{\mu}(x^{\alpha})$ and ${\bf b}(x^{\alpha})$. More specifically, the fluid velocity and temperature parameters are expanded as
\begin{equation}\label{linearization}
u_{\mu}(x^{\alpha})=\left(-1,\epsilon \beta_i(x^{\alpha})\right)+ \mathcal{O}(\epsilon^2),~~~{\bf b}(x^\alpha)={\bf b}_0+\epsilon {\bf b}_1(x^\alpha) + \mathcal{O}(\epsilon^2),
\end{equation}
where, as in~\cite{0712.2456}, we multiply $\beta_i(x^{\alpha})$ and ${\bf b}_1(x^{\alpha})$ by a small number $\epsilon$, which is an order counting parameter.
Below we are going to systematically trace only the terms linear in $\epsilon$ and set $\epsilon$ to one in the final expression of the fluid stress tensor.
${\bf b}_0$ corresponds to the equilibrium temperature of the dual fluid system while ${\bf b}_1(x^{\alpha})$ accounts for the dissipative corrections. In what follows, we use conformal symmetry to set ${\bf b}_0$ to one.

Substituting~(\ref{linearization}) into~(\ref{boosted bh}), the ``seed'' metric, i.e., a linearized version of~(\ref{metric1}) is
\begin{equation}\label{line element}
\begin{split}
ds^2=&2drdv-r^2f(r)dv^2+r^2d{\vec{x}}^2\\
&-\epsilon\left[2\beta_i(x^{\alpha}) drdx^i+\frac{2}{r^2}\beta_i(x^{\alpha}) dvdx^i+ \frac{4}{r^2} {\bf b}_1(x^{\alpha})dv^2\right]+\mathcal{O}(\epsilon^2),
\end{split}
\end{equation}
where the first line is the line element of $AdS_5$ black hole written in the ingoing Eddington-Finkelstein coordinate. As has been explained above, the metric~(\ref{line element}) does not solve the Einstein equations~(\ref{einstein eq}) at order $\epsilon$.
The term linear in $\epsilon$  in ~(\ref{line element}) is only a part of the metric we are after, and additional corrections at this
order have to be introduced. The full metric is formally written as
\begin{equation}
g=g^{(0)}+\epsilon g^{(1)}+\mathcal{O}(\epsilon^2),~~~\text{with}~~~
g^{(1)}=g_{\text{in}}^{(1)}+ g_{\text{corr}}^{(1)},
\end{equation}
where $g^{(0)}$ is the first line of~(\ref{line element}) and $g_{\text{in}}^{(1)}$ corresponds to the term linear in $\epsilon$ of~(\ref{line element}). The term $g_{\text{corr}}^{(1)}$ is the added correction, whose form has to be determined via solving the bulk Einstein equations~(\ref{einstein eq}).

In order to proceed with the computations, we fix gauge. Following~\cite{0712.2456}, we work in the ``background field'' gauge,
\begin{equation}\label{bk gauge}
g_{rr}=0,~~~g_{r\mu}\propto u_{\mu},~~~\textrm{Tr}\left[(g^{(0)})^{-1}g^{(1)}\right]=0.
\end{equation}
We pause to explain the results of the above gauge condition. The most general form of the undetermined metric correction $g_{\text{corr}}^{(1)}$ could be parameterized as,
\begin{equation}\label{undetermined metric}
ds_{\text{corr}}^2=\epsilon\left(\mathfrak{g}_{rr}dr^2+2\mathfrak{g}_{rv}drdv+ 2\mathfrak{g}_{ri}drdx^i+\mathfrak{g}_{vv}dv^2 +2\mathfrak{g}_{vi}dvdx^i+ \mathfrak{g}_{ij}dx^idx^j\right).
\end{equation}
The condition $g_{rr}=0$ implies $\mathfrak{g}_{rr}=0$. The metric components $g_{r\mu}$ can be read off from eqs.~(\ref{line element}) and~(\ref{undetermined metric}),
\begin{equation}
g_{r\mu}=\left(1,-\epsilon \beta_{i}\right)+\epsilon\left(\mathfrak{g}_{rv}, \mathfrak{g}_{ri}\right).
\end{equation}
Therefore, the second condition $g_{r\mu}\propto u_{\mu}$ amounts to requiring that
\begin{equation}
\left(\mathfrak{g}_{rv}, \mathfrak{g}_{ri}\right)\propto(1,-\epsilon \beta_i)~ \Longrightarrow \mathfrak{g}_{ri}=-\epsilon \beta_i \mathfrak{g}_{rv}.
\end{equation}
In other words, up to $\mathcal{O}(\epsilon)$, the vector component $\mathfrak{g}_{ri}$ will be set to zero. The last condition in (\ref{bk gauge}) gives a constraint,
\begin{equation}
\mathfrak{g}_{rv}+\frac{1}{2r^2}\sum_{i=1}^{3}\mathfrak{g}_{ii}=0.
\end{equation}
Under the gauge condition~(\ref{bk gauge}), the line element~(\ref{undetermined metric}) for $g_{\text{corr}}^{(1)}$ can be rewritten in the similar fashion as that of~\cite{0712.2456},
\begin{equation}\label{line element correction}
ds_{\text{corr}}^2=\epsilon\left(-3hdrdv+\frac{k}{r^2}dv^2 +r^2 h d{\vec{x}}^2+\frac{2}{r^2}j_{i}dvdx^i+r^2\alpha_{ij}dx^idx^j\right),
\end{equation}
where the trace part of $\mathfrak{g}_{ij}$ is explicitly denoted as scalar function $h$. Therefore, $\alpha_{ij}$ is a symmetric traceless tensor of rank two. Notice that all the metric components $\{h,~k,~j_{i},~\alpha_{ij}\}$ are functions of the bulk coordinates $\{x^{\alpha},r\}$. We shall find that these functions are functionals of the fluid velocity $\beta_i$, which we leave as an undetermined parameter. Their precise forms have to be determined by solving the bulk Einstein equations~(\ref{einstein eq}), supplemented with appropriate boundary conditions to be discussed in subsection~\ref{subsection22}.

Once the dual metric is found, the fluid stress tensor of the boundary CFT can be computed via holographic dictionary~\cite{hep-th/9802109,hep-th/9802150}. The dual fluid system is defined on the $r=\infty$ hypersurface. However, in an asymptotically AdS spacetime, holographic renormalization is needed to remove divergences near conformal boundary $r=\infty$. To proceed, we consider a hypersurface $\Sigma$ at constant $r$. The outgoing normal vector $n_{M}$ to $\Sigma$ is
\begin{equation}
n_{M}=\frac{\nabla_{M}r}{\sqrt{g^{MN}\nabla_{M}r\nabla_{N}r}},~~~~\textrm{and} ~~~~n^{M}=g^{MN}n_{N}.
\end{equation}
The induced metrics $\gamma_{MN}$ and $\gamma^{MN}$ on the hypersurface $\Sigma$ are constructed as
\begin{equation}
\gamma_{MN}=g_{MN}-n_{M}n_{N},~~~~\gamma^{MN}=g^{MN}-n^{M}n^{N}.
\end{equation}
The extrinsic curvature tensor $\mathcal{K}_{MN}$ of the hypersurface $\Sigma$ is
\begin{equation}
\mathcal{K}_{MN}=\frac{1}{2}\left(n^{A}\partial_{A}\gamma_{MN}+\gamma_{MA}\partial_N n^A + \gamma_{NA}\partial_{M}n^A\right).
\end{equation}
Using the formula of~\cite{hep-th/9902121,hep-th/9806087}, the stress tensor for the dual fluid is
\begin{equation}\label{stress tensor1}
T_{\nu}^{\mu}=\lim_{r\rightarrow \infty}\tilde{T}_{\nu}^{\mu}(r)=-2\lim_{r\rightarrow \infty}r^4\left(\mathcal{K}_{\nu}^{\mu}-\mathcal{K}\gamma_{\nu}^{\mu}+ 3\gamma_{\nu}^{\mu} -\frac{1}{2}G^{\mu}_{\nu}\right),
\end{equation}
where $G^{\mu}_{\nu}$ is the Einstein tensor constructed from the induced metric $\gamma_{\mu\nu}$ and $\mathcal{K}\equiv \gamma^{\mu\nu}\mathcal{K}_{\mu\nu}$. The last two terms in eq.~(\ref{stress tensor1}) are the counter-terms needed to remove divergences near the conformal boundary $r= \infty$.

Applied to the metric~(\ref{line element correction}), the fluid stress tensor can be expressed in terms of the functions $\{h,~k,~j_{i},~\alpha_{ij}\}$. We here record all the components of $\tilde{T}^{\mu}_{\nu}$,
\begin{equation}\label{stress tensor}
\left\{
\begin{aligned}
\tilde{T}_{0}^{0}=&-3(1-4\epsilon {\bf b}_1)+\frac{\epsilon}{2r}\left\{-6rk+4r^4\partial \beta- 4\partial j-r^3\partial_i \partial_j\alpha_{ij}+18(r^5-r)h\right.\\
&\left.+6(r^6-r^2)\partial_rh+2r^3\partial^2h+6r^4\partial_vh\right\},\\
\tilde{T}^{0}_{i}=&\frac{\epsilon}{2r^4}\left\{2\left[4r^4\beta_i-4(r^4-1) j_i + r^7\partial_v \beta_i-r^3\partial_i k+(r^5-r)\partial_r j_i\right]\right.\\
&\left.-r^2\left(-\partial^2 j_i + \partial_i\partial j + r^4\partial_v\partial_k\alpha_{ik}-2r^4\partial_v\partial_ih-3r^5\partial_ih\right)\right\},\\
\tilde{T}^{i}_{0}=&-\frac{\epsilon}{2r^3}\left\{2\left[4r^3\beta_i-4r^3 j_i +r^6\partial_v \beta_i -r^2\partial_i k +(r^4-1) \partial_r j_i\right]\right.\\
&\left.+r\left[\partial^2 j_i-\partial_i\partial j-r^4\partial_v\partial_k\alpha_{ik}- 2r^4\partial_v\partial_ih-3(r^6-r^2)\partial_ih\right]\right\},\\
\tilde{T}_{j}^{i}=&\delta^{i}_{j}(1-4\epsilon {\bf b}_1)+\frac{\epsilon}{2r^4}\delta^i_j \left\{r^2\left[-\partial^2 k+(1-r^4)\partial_k\partial_l \alpha_{kl} +2\partial_v\partial j\right]\right.\\
&\left.-2\left[(1-r^4)k-2r^7\partial\beta+2r^3\partial j-r^3 \partial_v k +(r^5-r)\partial_r k \right]+r^6\partial^2h\right.\\
&\left.-2r^6\partial_v^2h+2\left[\left(3-12r^4+9r^5\right)h+(r^3-r^7)\partial_vh +(2r-4r^5+2r^9)\partial_rh\right]\right\}\\
&+\frac{\epsilon}{2r^2}\left\{-2r\left[2r^4\partial_{(i}\beta_{j)}-2\partial_{(i}j_{j)} +r^4\partial_v\alpha_{ij}+(r^6-r^2)\partial_r\alpha_{ij}\right]-r^4\partial_i\partial_j h\right.\\
&\left.+\left[\partial_i\partial_j k+(1-r^4)\partial^2\alpha_{ij} +2(r^4-1) \partial_{k}\partial_{(i}\alpha_{j)k}-2\partial_v\partial_{(i}j_{j)}+r^4\partial_v^2 \alpha_{ij}\right]\right\},
\end{aligned}
\right.
\end{equation}
where the notation $\partial_{(i}\beta_{j)}$ etc stands for symmetrization over the indices $i$ and $j$. It is important to notice that in the above expression for $\tilde T^\mu_\nu$, we have already dropped  terms, which explicitly vanish as $r\to \infty$.

\subsection{Dynamics: the bulk gravity and the boundary hydrodynamics} \label{subsection22}
In this subsection we write down dynamical equations which determine the functions $\{h,~k,~j_i,~\alpha_{ij}\}$ in the bulk. Having these functions at hand, we extract the fluid stress tensor via eq.~(\ref{stress tensor}). To proceed, we have to specify proper boundary conditions for these metric functions. The first one is a regularity requirement for the metric over the whole range of $r$, in particular at unperturbed horizon $r=1$. This follows from our choice of the ingoing Eddington-Finkelstein coordinate in which the metric is free of any coordinate singularity. The second boundary condition comes from
asymptotic considerations. In the present paper we restrict our analysis to boundary fluid dynamics in \emph{flat} spacetime with the Minkowski metric $\eta_{\mu\nu}$. Therefore, we require that the metric correction does not modify the asymptotic features of the metric~(\ref{metric1}). This condition tightly constrains the large $r$ behavior for metric functions $\{h,~k,~j_i,~\alpha_{ij}\}$. Specifically, as $r\to \infty$, their behaviors should be restricted as
\begin{equation}\label{AdS constraint}
h< \mathcal{O}(r^0),~~~k<\mathcal{O}(r^4),~~~j_i<\mathcal{O}(r^4),~~~ \alpha_{ij}<\mathcal{O}(r^0).
\end{equation}
Yet some integration constants remain unfixed due to a freedom of defining fluid velocity. We follow~\cite{0712.2456} and choose a frame for the dual fluid system. We will work in the ``Landau frame'' defined by
\begin{equation}\label{frame convention}
u_{\mu}T^{\mu\nu}_{\text{Diss}}=0.
\end{equation}

We are now in the position to study dynamics of the bulk gravity. As pointed out in  section $3.2$ of~\cite{0712.2456}, there is one redundancy among the total $15$ components of the Einstein equations~(\ref{einstein eq}). Similarly to~\cite{0712.2456}, we classify the remaining $14$ equations into constraint equations and dynamical ones. In order to construct the fluid stress tensor, we only solve the dynamical equations. This will lead us to an ``off-shell'' fluid stress tensor with undetermined fluid velocity  but with the transport coefficient functions fixed. The constraint equations will be  shown to be equivalent to the conservation law of thus constructed fluid stress tensor.

The first equation we are to consider is $E_{rr}=0$,
\begin{equation}\label{h eq}
5\partial_rh+r\partial_r^2h=0.
\end{equation}
A generic solution to $h$ is
\begin{equation}
h(x^{\alpha},r)=s_0(x^{\alpha})+\frac{s_1(x^{\alpha})}{r^4},
\end{equation}
where $s_0$ and $s_{1}$ are arbitrary functions of the boundary coordinates $x^{\alpha}$. A nonzero function $s_0$ would violate the asymptotic requirement for $h$ as specified in eq.~(\ref{AdS constraint}). In addition, $s_1\neq0$ will result in $T_{\text{Diss}}^{00}\neq 0$ as can be seen from eq.~(\ref{stress tensor}). Therefore, the constraint on the asymptotic behavior at the infinity and the ``Landau frame'' convention enforce $h=0$.

We proceed by considering $E_{rv}=0$,
\begin{equation}\label{k eq}
3r^2\partial_r k=6r^4\partial\beta+r^3\partial_v\partial\beta-2\partial j- r\partial_r\partial j-r^3\partial_i\partial_j\alpha_{ij}.
\end{equation}
Clearly, the function $k$ cannot be found until the vector $j_i$ and tensor $\alpha_{ij}$ are computed and we will postpone integration of eq.~(\ref{k eq}) until first solving for $j_i$ and $\alpha_{ij}$. Fortunately, the dynamical equations for $j_i$ and $\alpha_{ij}$ can be disentangled from $k$.

The dynamical equation for $j_i$ follows from $E_{ri}=0$,
\begin{equation}\label{ji eq}
-\partial_r^2 j_i=\left(\partial^2\beta_i-\partial_i\partial\beta\right)+3r \partial_v \beta_i-\frac{3}{r}\partial_r j_i +r^2\partial_r\partial_j\alpha_{ij}.
\end{equation}
The diagonal and off-diagonal components of $\alpha_{ij}$ should be treated separately. We first consider the off-diagonal components emerging from $E_{ij}=0$ with $i\neq j$,
\begin{equation}\label{non-diagonal eq}
\begin{split}
0=&(r^7-r^3)\partial_r^2 \alpha_{ij}+(5r^6-r^2)\partial_r\alpha_{ij}+
2r^5\partial_v\partial_r\alpha_{ij}+3r^4\partial_v\alpha_{ij}\\
&+r^3\left\{\partial^2\alpha_{ij}-\left(\partial_i\partial_k\alpha_{jk} +
\partial_j\partial_k\alpha_{ik}\right)\right\}+\left(\partial_i j_j+\partial_j j_i\right)-r\partial_r\left(\partial_i j_j+\partial_j j_i\right)\\
&+3r^4\left(\partial_i\beta_j+\partial_j\beta_i\right)+
r^3\partial_v\left(\partial_i\beta_j+\partial_j\beta_i\right).
\end{split}
\end{equation}
The diagonal components of $\alpha_{ij}$ are coupled with the function $k$. We present the dynamical equation for $\alpha_{11}$,
\begin{equation}\label{alpha11p}
\begin{split}
0=&(r^4-1)\partial_r^2\alpha_{11}+\frac{5r^4-1}{r}\partial_r \alpha_{11}+
2r^2\partial_v\partial_r\alpha_{11}+3r\partial_v\alpha_{11}\\
&-\left(\partial_3^2\alpha_{22}-2\partial_2\partial_3\alpha_{23}+\partial_2^2\alpha_{33}\right) -\frac{2}{r^3}\left(\partial j-\partial_1 j_1\right)+ \frac{2}{r^2}\partial_r\left(\partial j-\partial_1 j_1\right)\\
&-6r(\partial\beta-\partial_1\beta_1)-2\partial_v\left(\partial\beta-\partial_1\beta_1\right) + \partial_r^2k,
\end{split}
\end{equation}
where the term $\partial_r^2k$ will be eliminated using eqs.~(\ref{k eq}) and~(\ref{ji eq}),
\begin{equation}
\partial_r^2 k=4r\partial\beta+\frac{4}{3}\partial_v\partial\beta + \frac{4}{3r^3} \partial j -\frac{4}{3r^2}\partial_r\partial j-\frac{1}{3}\partial_i\partial_j \alpha_{ij}.
\end{equation}
The equation for $\alpha_{11}$ can be put into a new form,
\begin{equation}\label{diagonal eq}
\begin{split}
0=&(r^7-r^3)\partial_r^2\alpha_{11}+(5r^6-r^2)\partial_r\alpha_{11}+ 2r^5 \partial_v\partial_r \alpha_{11} + 3r^4\partial_v\alpha_{11} \\
&-r^3\left(\partial_3^2\alpha_{22}-2\partial_2\partial_3\alpha_{23} + \partial_2^2\alpha_{33}+\frac{1}{3}\partial_i\partial_j \alpha_{ij}\right)\\
&+\left(\partial_1 j_1+\partial_1 j_1-\frac{2}{3}\partial j\right)-r\partial_r \left(\partial_1 j_1+\partial_1 j_1-\frac{2}{3}\partial j\right)\\
&+3r^4\left(\partial_1\beta_1+\partial_1\beta_1-\frac{2}{3}\partial\beta\right)+ r^3 \partial_v \left(\partial_1\beta_1+\partial_1\beta_1-\frac{2}{3}\partial\beta\right).
\end{split}
\end{equation}
Similar equations hold for $\alpha_{22}$ and $\alpha_{33}$. Remarkably, the equations for the off- and diagonal components of $\alpha_{ij}$ can be combined into a unified form,
\begin{equation}\label{alphaij eq}
\begin{split}
0=&(r^7-r^3)\partial_r^2 \alpha_{ij}+(5r^6-r^2)\partial_r\alpha_{ij}+
2r^5\partial_v\partial_r\alpha_{ij}+3r^4\partial_v\alpha_{ij}\\
&+r^3\left\{\partial^2\alpha_{ij}-\left(\partial_i\partial_k\alpha_{jk} +
\partial_j\partial_k\alpha_{ik}-\frac{2}{3}\delta_{ij} \partial_k\partial_l\alpha_{kl}\right)\right\}\\
&+\left(\partial_i j_j+\partial_j j_i-\frac{2}{3}\delta_{ij} \partial j\right)-r\partial_r\left(\partial_i j_j+\partial_j j_i-\frac{2}{3}\delta_{ij} \partial j\right)\\
&+3r^4\left(\partial_i\beta_j+\partial_j\beta_i-\frac{2}{3}\delta_{ij} \partial\beta\right)+
r^3\partial_v\left(\partial_i\beta_j+\partial_j\beta_i-\frac{2}{3}\delta_{ij} \partial\beta\right).
\end{split}
\end{equation}

We are to solve these second order partial differential equations~(\ref{ji eq}) and~(\ref{alphaij eq}). As has been outlined above, $j_i$ and $\alpha_{ij}$ are linear functionals of $\beta_i$. They can be uniquely decomposed as
\begin{equation}\label{decomposition}
\left\{
\begin{aligned}
j_i=&a(\omega,q,r)\beta_i+b(\omega,q,r)\partial_i\partial\beta,\\
\alpha_{ij}=&2c(\omega,q,r)\sigma_{ij}+ d(\omega,q,r)\pi_{ij},
\end{aligned}
\right.
\end{equation}
where $\sigma_{ij}$ and $\pi_{ij}$ are defined in~(\ref{sigma}) and~(\ref{tensor structures}). The above decomposition is obviously inspired by the special structure of the source terms in~(\ref{ji eq}) and~(\ref{alphaij eq}), which are composed of $\beta_i$ and its derivatives only. In writing down~(\ref{decomposition}), we ignored the homogeneous part of the general solutions for~(\ref{ji eq}) and~(\ref{alphaij eq}), as they would modify the definition of fluid velocity~(\ref{frame convention}) and the boundary requirement~(\ref{AdS constraint}). In addition, the possible basis vector and tensor constructed from the temperature variation ${\bf b}_1$ do not appear in~(\ref{decomposition}). One would add such structures in~(\ref{decomposition}) with associated coefficient functions. Then, these new coefficient functions would obey \emph{homogeneous} differential equations. The boundary conditions summarized in subsection~\ref{subsection22}, in particular the ``Landau frame'' convention, force these new coefficient functions to trivially vanish. Indeed, the generalized Navier-Stokes equations~(\ref{NS eqs}) relate temperature gradient to derivatives of fluid velocity. Therefore, the derivatives of ${\bf b}_1$ should not be treated as linearly independent blocks in~(\ref{decomposition}).

Eqs.~(\ref{ji eq}) and~(\ref{alphaij eq}) get converted into dynamical equations for the functions $a,b,c,d$
\begin{equation}\label{abcd eqs}
\left\{
\begin{aligned}
0=&r\partial_r^2a-3\partial_ra-q^2r^3\partial_r c-3 i \omega r^2-q^2 r,\\
0=&r\partial_r^2b-3\partial_r b+ \frac{1}{3}r^3\partial_r c -\frac{2}{3} r^3 q^2\partial_r d -r,\\
0=&(r^7-r^3)\partial_r^2 c+(5r^6-r^2)\partial_r c - 2i \omega r^5\partial_r c\\
&-r\partial_r a + a- 3i \omega r^4 c+ 3r^4- i\omega r^3,\\
0=&(r^7-r^3)\partial_r^2 d+(5r^6-r^2)\partial_r d-2i \omega r^5\partial_r d\\
&+\frac{q^2}{3} r^3d-3i\omega r^4 d + 2b-2r\partial_r b-\frac{2}{3}r^3 c.
\end{aligned}
\right.
\end{equation}
The dynamical equation~(\ref{k eq}) is rendered into the form,
\begin{equation}\label{keq momentum}
\partial_r k=\left\{2r^2-\frac{1}{3}i\omega r-\frac{2}{3r^2}(a-q^2 b)-\frac{1}{3r}\left(\partial_r a -q^2 \partial_r b\right) -\frac{q^2r}{9} \left(-4c + 2q^2 d\right)\right\}\partial\beta.
\end{equation}
The replacement rule $\partial_{\mu}\to\left(-i\omega,i\vec{q}\right)$ is implied in eqs.~(\ref{decomposition}),~(\ref{abcd eqs}) and~(\ref{keq momentum}). The equations (\ref{abcd eqs}) are the holographic RG-flow type equations for the viscosity functions. Together with corresponding solutions, these equations constitute our main results of this paper. Through the relations~(\ref{decomposition}) and~(\ref{stress tensor}), the fluid stress tensor of the boundary CFT is determined by asymptotic behaviors of the functions $a,b,c,d$ as $r\to\infty$. In section~\ref{section4} we will study these asymptotic behaviors by exploring the equations (\ref{abcd eqs}) near the boundary.  Here we summarize these studies,
\begin{equation}\label{abcd asymp}
\begin{split}
a(\omega,q,r)&=-i\omega r^3+\mathcal{O}\left(\frac{1}{r}\right), ~~~~~~~~~~~~~b(\omega,q,r)=-\frac{1}{3}r^2+\mathcal{O}\left(\frac{1}{r}\right),\\
c(\omega,q,r)&=\frac{1}{r}-\frac{\eta(\omega,q^2)}{8r^4}+\mathcal{O} \left(\frac{1}{r^5}\right),~~~d(\omega,q,r)=-\frac{\zeta(\omega,q^2)}{4r^4} + \mathcal{O}\left(\frac{1}{r^5}\right),
\end{split}
\end{equation}
where the momenta-dependent functions $\eta(\omega,q^2)$ and $\zeta(\omega,q^2)$ are formally introduced as coefficients in the asymptotic expansion. The boundary conditions~(\ref{AdS constraint}) and~(\ref{frame convention}) have been already applied while deriving eq.~(\ref{abcd asymp}). Asymptotic behaviors as $r\to\infty$ by themselves cannot fix the coefficients $\eta(\omega,q^2)$ and $\zeta(\omega,q^2)$. To find them, we have to solve the RG equations (\ref{abcd eqs}) in full starting from the horizon and integrating up to the boundary. The regularity requirement at $r=1$ is found to be
sufficient to fix $\eta$ and $\zeta$ uniquely. We postpone completing this computation until sections~\ref{section3} (analytic) and~\ref{section4} (numeric).

Meanwhile, based on (\ref{abcd asymp}), as $r\to\infty$ the components $j_i$ and $\alpha_{ij}$ behave as
\begin{equation} \label{large r}
\left\{
\begin{aligned}
j_{i}\to&-i\omega r^3\beta_i-\frac{1}{3}r^2\partial_i\partial\beta + \mathcal{O}\left(\frac{1}{r}\right),\\
\alpha_{ij}\to&\left(\frac{2}{r}-\frac{\eta(\omega,q^2)}{4r^4}\right)\sigma_{ij} -\frac{\zeta(\omega,q^2)}{4r^4}\pi_{ij}+ \mathcal{O}\left(\frac{1}{r^5}\right).
\end{aligned} \right.
\end{equation}
The large $r$ behavior of the function $k$ follows from first integrating~(\ref{keq momentum}) over $r$ and then making use of eq.~(\ref{abcd asymp}),
\begin{equation}\label{k large r}
k\to\frac{2}{3}\left(r^3+i\omega r^2\right)\partial\beta+ \mathcal{O}\left(\frac{1}{r^2}\right),~~~\text{as}~~~r\to\infty,
\end{equation}
where the integration constant is fixed by the ``Landau frame'' convention~(\ref{frame convention}).

Substituting eqs.~(\ref{large r}) and~(\ref{k large r}) into eqs.~(\ref{stress tensor}) and~(\ref{stress tensor1}), we derive the fluid stress tensor as summarized in eqs.~(\ref{ideal stress tensor}) and~(\ref{stress tensor nonperturb}). This establishes, fully and uniquely, the relation~(\ref{stress tensor nonperturb}) between the dissipative part of the stress tensor and the large $r$ behaviors of the functions $c$ and $d$, as encoded in the viscosity functions $\eta(\omega,q^2)$ and $\zeta(\omega,q^2)$.
\subsection{Generalized Navier-Stokes equations and spectrum of small fluctuations}\label{subsection23}
Generalized, all order Navier-Stokes equations follow from the bulk constraints $E_{vv}=E_{vi}=0$. We find it more convenient to study suitable linear combinations of these constraints and dynamical equations. More specifically, the combination $r^2f(r)E_{vr}+E_{vv}=0$ states
\begin{equation}\label{constraint1}
\begin{split}
12\partial_v {\bf b}_1=&4\partial\beta+\frac{4}{r}\partial^2 {\bf b}_1-\left(\frac{1}{r}+r^3\right) \partial_v\partial\beta-4\partial j\\
&-\frac{1}{r}\partial^2k+ 3\partial_v k+ \frac{2}{r} \partial_v \partial j- \left(\frac{1}{r^3}-r\right) \partial_r \partial j.
\end{split}
\end{equation}
The second constraint $r^2f(r)E_{ri}+E_{vi}=0$ yields
\begin{equation}\label{constraint2}
\begin{split}
4\partial_i {\bf b}_1=&4\partial_v\beta_i+r^3\partial_i\partial\beta-r^3\partial^2\beta_i +r^3 \partial_v^2\beta_i+\frac{1}{r}\partial^2 j_i-\frac{1}{r}\partial_i\partial j+\partial_i k \\
&-4\partial_v j_i-r^3\partial_v\partial_j\alpha_{ij}+(r-r^5)\partial_r\partial_j \alpha_{ij}- r\partial_r\partial_i k+r\partial_r\partial_v j_i.
\end{split}
\end{equation}
Taking the large $r$ limit of~(\ref{constraint1}) and~(\ref{constraint2}) results in
\begin{equation}\label{NS eqs}
\partial_v {\bf b}_1=\frac{1}{3}\partial\beta,~~~~~~~
\partial_i {\bf b}_1=\partial_v\beta_i-\frac{\eta(\partial_v,\partial^2)}{24}\left(\partial_i\partial \beta+ 3\partial^2\beta_i\right)-\frac{\zeta(\partial_v,\partial^2)}{6}\partial^2 \partial_i\partial\beta.
\end{equation}
Eqs.~(\ref{NS eqs}) are equivalent to the boundary stress tensor conservation law $\partial_{\mu}T^{\mu\nu}=0$, which determines the temperature and velocity profiles as  functions of time, provided initial configuration is specified.

We close this section by studying the spectrum of small fluctuations of the local fluid mechanical variables. We consider a plane wave ansatz for the velocity
 $\beta_i(x^{\alpha})$ and temperature ${\bf b}(x^{\alpha})$,
\begin{equation}\label{plane wave}
\beta_{i}(x^{\alpha})=\delta\beta_i e^{-i\omega v+iq_jx^j},~~~~~~~~
{\bf b}(x^{\alpha})=1+\delta {\bf b}_1 e^{-i\omega v+iq_jx^j}.
\end{equation}
Substituting this ansatz into eqs.~(\ref{NS eqs}) results in a set of four homogeneous linear equations in the amplitudes $\delta\beta_i$ and $\delta {\bf b}_1$. Coefficients of these equations are functions of $\omega$ and $q_i$. These equations have nontrivial solutions if and only if the matrix formed out of these coefficient functions has zero determinant. For transverse case, where $\vec{q}\perp \vec{\beta}$, we are led to  dispersion relation for shear mode,
\begin{equation}\label{shear mode}
\omega+\frac{i}{8}\eta(\omega,q^2)q^2=0.
\end{equation}
Similarly, we obtain dispersion relation for sound mode by taking $\vec{q}\parallel\vec{\beta}$,
\begin{equation}\label{sound mode}
q^2-3\omega^2-\frac{i}{2}\eta(\omega,q^2)\omega q^2+\frac{i}{2}\zeta(\omega,q^2)\omega q^4=0.
\end{equation}
Notice that the second viscosity function $\zeta(\omega,q^2)$ does not contribute to the shear mode.
The dispersion relations (\ref{shear mode},\ref{sound mode}),  are exact, namely valid for any value of $\omega$ and $q$. As has been already discussed in~\cite{0905.4069}, beyond small $\omega$, $q$ limit, these equations generate infinitely many solutions, which are the quasinormal modes of the dual theory.

In the next section, we will determine the viscosity functions  $\eta(\omega,q^2)$ and $\zeta(\omega,q^2)$ in the hydrodynamic limit $\omega,~q_i\ll 1$.
The results have been quoted in eq.~(\ref{viscosity funs}). With these expressions at hand, we consider corrections to dispersion relations due to the higher order derivative terms. Solving the dispersion equations~(\ref{shear mode}) and~(\ref{sound mode}) perturbatively, we obtain
\begin{equation}
\begin{split}
\text{shear wave:}~~\omega=&-\frac{i}{4}q^2-\frac{i}{32}(1-\ln{2})q^4+\cdots,\\
\text{sound wave:}~~\omega=&\pm\frac{1}{\sqrt{3}}q-\frac{i}{6}q^2\pm\frac{1}{24\sqrt{3}} \left(3-2\ln{2}\right)q^3\\
&+\frac{i}{288}\left(8-\frac{\pi^2}{3}+4\ln^2{2}-4\ln{2}\right)q^4+\cdots.
\end{split}
\end{equation}
These results can be compared with those obtained in~\cite{0712.2451,0712.2456}. The shear mode dispersion relation agrees with the quasi-normal mode computations of~\cite{0712.2451} up to $q^4$ terms. This further validates the argument of~\cite{0712.2451,0712.2456} that the third order derivative terms in the fluid stress tensor are crucial in correctly producing the shear wave mode at order $q^4$. However, the sound wave dispersion relation presented here is different from the one obtained in~\cite{0712.2451,0712.2456}: it is being corrected by the third order derivative in the fluid stress tensor. The analytical expressions for the dispersion relations should agree at small $\omega,q_i$ with numerical results of~\cite{hep-th/0506184}.

\section{Viscosity functions in the hydrodynamic regime: Perturbative analysis}\label{section3}
In this section we solve eqs.~(\ref{abcd eqs}) and~(\ref{keq momentum}) perturbatively in the hydrodynamic regime $\omega,~q_i\ll 1$. We then compute the fluid stress tensor using thus constructed perturbative metric correction, up to third order in the derivative expansion. Up to second order, we reproduce some well-known results from the literature, validating correctness of our formalism. We also compute new third order transport coefficients. Finally, we give a formal construction to any order in the derivative expansion, which is fully consistent with the numerical analysis of section~\ref{section4}.

To trace the order in the derivative expansion, we multiply $\omega$ and $q_i$ by a small parameter $\lambda$
\begin{equation}
\omega\rightarrow \lambda\omega,~~~~q_i\rightarrow \lambda q_i.
\end{equation}
The functions $a,~b,~c,~d$ are then expanded in powers of $\lambda$,
\begin{equation}\label{lambda expansion}
\begin{split}
a(\omega,q,r)&=\sum_{n=0}^{\infty}\lambda^{n} a_n(\omega,q,r),~~~ b(\omega,q,r)=\sum_{n=0}^{\infty}\lambda^{n} b_n(\omega,q,r),\\
c(\omega,q,r)&=\sum_{n=0}^{\infty}\lambda^{n} c_n(\omega,q,r),~~~
d(\omega,q,r)=\sum_{n=0}^{\infty}\lambda^{n} d_n(\omega,q,r).
\end{split}
\end{equation}
Correspondingly, the metric correction is counted by powers of $\lambda$,
\begin{equation}
k=\sum_{n=1}^{\infty}k^{(n)},~~~j_i=\sum_{n=1}^{\infty}j_i^{(n)},~~~\alpha_{ij}= \sum_{n=1}^{\infty}\alpha_{ij}^{(n)},
\end{equation}
which follows from~(\ref{lambda expansion}). Notice, however, that different orders in the coefficients $a_n,\cdots,d_n$ mix due to explicit derivatives in the decomposition (\ref{decomposition}). It is straightforward to write down equations for $a_n,\cdots,d_n$ at each order in $\lambda$. In what follows, we explicitly solve these equations imposing the boundary conditions discussed in section~\ref{section2}.

\subsection{Perturbative solution to the metric correction} \label{subsection31}
\paragraph{Metric correction at zeroth order.}
To the lowest order, we have equation for $a_0$ only,
\begin{equation}
r\partial_r^2 a_0-3 \partial_r a_0=0,
\end{equation}
whose generic solution is
\begin{equation}\label{rute a0}
a_0(\omega,q,r)=Cr^4+C^{\prime}.
\end{equation}
In the above solution, the constant $C$ multiplies a non-normalizable mode, which deforms the metric of the boundary field theory. It has to vanish by the condition~(\ref{AdS constraint}). The remaining constant $C^{\prime}$ corresponds to a shift of the fluid velocity and is  set to zero by the ``Landau frame'' convention~(\ref{frame convention}). Therefore, there exists no nontrivial solution for $a_0$, which is consistent with  intuition that metric correction appears starting from first order in the derivative expansion.

\paragraph{Metric correction at first order.}
Up to the first order in the derivative expansion, it is sufficient to consider the system of differential equations,
\begin{equation} \label{1st order}
\left\{
\begin{aligned}
&0=(r^7-r^3)\partial_r^2 c_0 +(5r^6-r^2)\partial_r c_0 +3r^4,\\
&0=r\partial_r^2 a_1-3 \partial_r a_1-3i\omega r^2,
\end{aligned}
\right.
\end{equation}
where $a_0=0$ was already used. We study~(\ref{1st order}) as an example of how to fix integration constants. We first rewrite equation for $c_0$ as
\begin{equation}
\begin{split}\label{c0 eq}
(r^7-r^3)\partial_r^2 c_0 +(5r^6-r^2)\partial_r c_0 +3r^4=0\\
\Longrightarrow~r^2\partial_r\left[(r^5-r)\partial_r c_0\right]+3r^4=0.
\end{split}
\end{equation}
First integration is done from $1$ to $r$:
\begin{equation}\label{c0 eq 1st}
\partial_r c_0=\frac{1}{r^5-r}\int_{1}^{r}(-3y^2)dy + \frac{\sharp_{c_0}}{r^5-r},
\end{equation}
where the integration constant $\sharp_{c_0}$ is set to zero by the regularity condition at $r=1$. Then, as $r\to\infty$, the right-handed side of~(\ref{c0 eq 1st}) falls off like $\sim 1/r^2$, so that it is valid to integrate the above equation from $r$ to $\infty$,
\begin{equation}
\begin{split}\label{2nd integral}
c_0(\omega,q,r)&=-\int_{r}^{\infty}\frac{dx}{x^5-x}\int_{1}^{x}(-3y^2)dy+\natural_{c_0}\\
&=\frac{1}{4}\left[\ln{\frac{(1+r^2)(1+r)^2}{r^4}}-2 \arctan{(r)}+\pi\right] +\natural_{c_0}\\
&\equiv F(r)+\natural_{c_0}.
\end{split}
\end{equation}
To fix the integration constant $\natural_{c_0}$, we consider the large $r$ behavior of $F(r)$,
\begin{equation}
F(r)\to \frac{1}{r}-\frac{1}{4r^4}+\mathcal{O}\left(\frac{1}{r^5} \right),~~~~\text{as}~r\rightarrow \infty.
\end{equation}
Therefore, to keep the asymptotic requirement for $\alpha_{ij}$ as given in~(\ref{AdS constraint}), the integration constant $\natural_{c_0}=0$. So,
\begin{equation}\label{c0 integral}
c_0(\omega,q,r)= F(r).
\end{equation}
A remark about the integration constant $\natural_{c_0}$ is worthy. In principle, the outer integral in eq.~(\ref{2nd integral}) might also be done from $1$ to $r$, but with a new integration constant different from $\natural_{c_0}$. This constant would have to be determined by the same asymptotic considerations. The final result for $c_0$ is still given by eq.~(\ref{c0 integral}).

We proceed with $a_1$,
\begin{equation}
\begin{split}
&r\partial_r^2 a_1-3 \partial_r a_1-3i\omega r^2=0\\
&\Longrightarrow r^4\partial_r(r^{-3}\partial_ra_1)=3i\omega r^2\\
&\Longrightarrow a_1(\omega,q,r)=-i\omega r^3+A(\omega,q^2)r^4+A^{\prime}(\omega,q^2).
\end{split}
\end{equation}
The functions $A(\omega,q^2)$ and $A^{\prime}(\omega,q^2)$ should equal zero, following the same arguments as below eq.~(\ref{rute a0}). We arrive at a first nontrivial expression for the function $a$,
\begin{equation}\label{a1}
a_1(\omega,q,r)=-i\omega r^3.
\end{equation}
Up to the first order in the derivative of the fluid velocity, equation~(\ref{keq momentum}) simplifies to
\begin{equation}
\partial_r k^{(1)}=2r^2\partial\beta,
\end{equation}
and the result for $k^{(1)}$ is
\begin{equation}\label{k(1)}
k^{(1)}=\frac{2}{3}r^3\partial\beta,
\end{equation}
where we use the convention~(\ref{frame convention}) to set the integration constant in~(\ref{k(1)}) to zero.

We summarize the metric correction at the first order in the derivative of the fluid velocity,
\begin{equation}\label{1st order correction}
k^{(1)}=\frac{2}{3}r^3\partial\beta,~~~j_i^{(1)}=r^3\partial_v \beta_i,~~~
\alpha_{ij}^{(1)}\xlongrightarrow{r\to\infty}\left(\frac{2}{r}-\frac{1}{2r^4}\right) \sigma_{ij}+\mathcal{O} \left(\frac{1}{r^5}\right).
\end{equation}

\paragraph{Metric correction at second order.}
The analysis above can be straightforwardly extended to the second order. The relevant equations are
\begin{equation} \label{2nd order}
\left\{
\begin{aligned}
&0=r\partial_r^2 b_0-3 \partial_r b_0 +\frac{1}{3}r^3 \partial_r c_0-r, \\
&0=(r^7-r^3)\partial_r^2 c_1 +(5r^6-r^2)\partial_r c_1-2i\omega r^5\partial_r c_0  -3i\omega r^4 c_0+i\omega r^3,\\
&0=r\partial_r^2 a_2-3 \partial_r a_2-q^2 r^3 \partial_r c_0- q^2 r,
\end{aligned}
\right.
\end{equation}
where the expression~(\ref{a1}) for $a_1$ has been used to simplify the equation for $c_1$.

The function $b_0$ obeys the same equation as $a_0$ except for the source term,
\begin{equation}
\partial_r\left(r^{-3}\partial_r b_0\right)=r^{-4}\left(-\frac{1}{3}r^3 \partial_r c_0 +r \right),
\end{equation}
where the large $r$ behavior for the source term is $\sim1/r^3$. The generic solution for $b_0$ can be obtained by integration over $r$,
\begin{equation}
b_0(\omega,q,r)=-\int_{1}^{r}dx~x^3\int_{x}^{\infty}\left(\frac{1}{y^3}-\frac{\partial_y c_0(y)}{3y} \right)dy+\sharp_{b_0}r^4+\natural_{b_0}.
\end{equation}
The outer integral is taken from $1$ to $r$ to make it well-defined. The integration constants $\sharp_{b_0}$ and $\natural_{b_0}$ will be again fixed by the asymptotic conditions and ``Landau frame'' convention as done for $a_0$,
\begin{equation}
\sharp_{b_0}=0,~~~\natural_{b_0}=-\frac{3}{8}.\nonumber
\end{equation}
The solution for $b_0$ now reads
\begin{equation}\label{b0 solution}
\begin{split}
b_0(\omega,q,r)&=-\int_{1}^{r}x^3dx\int_{x}^{\infty}\left[\frac{1}{y^3}-\frac{1}{3y} \partial_y c_0(y)\right]dy-\frac{3}{8},\\
&\longrightarrow-\frac{1}{3}r^2+\mathcal{O}\left(\frac{1}{r}\right), ~~~\text{as}~~~r\to\infty.
\end{split}
\end{equation}

The equation for $c_1$ can be solved similarly to its zeroth order counterpart $c_0$. We present  final results,
\begin{equation}
\begin{split}
&r^2\partial_r\left[(r^5-r)\partial_r c_1\right]=2i\omega r^5\partial_r c_0 -i\omega r^3 +3i\omega r^4 c_0,\\
&\Longrightarrow~c_1(\omega,q,r)=-i\omega \int_{r}^{\infty}\frac{dx}{x^5-x}\int_{1}^{x}dy \left[2y^3\partial_yc_0(y)-y+3y^2c_0(y)\right],
\end{split}
\end{equation}
which, as $r\to\infty$, falls off as
\begin{equation}
c_1(\omega,q,r\rightarrow \infty)= -\frac{i\omega}{4r^4}\left(1-\frac{\ln{2}}{2}\right)+ \mathcal{O}\left(\frac{1}{r^5}\right).
\end{equation}

Since the source term in the equation for $a_2$ decays rather rapidly in the large $r$ regime,  solution for $a_2$ is
\begin{equation}
a_2(\omega,q,r)=\int_r ^{\infty}x^3dx \int_x^{\infty}dy \left[\frac{q^2}{y} \partial_y c_0(y) +\frac{q^2}{y^3}\right]\xlongrightarrow{r\to\infty} \frac{q^2}{5r}+ \mathcal{O}\left(\frac{1}{r^2}\right),
\end{equation}
where the integration constants are fixed in a similar manner as in the above cases of $a_0$ and $a_1$.

The second order correction $k^{(2)}$ is solved by,
\begin{equation}
\begin{split}
&\partial_rk^{(2)}=\left\{-\frac{1}{3}i\omega r-\frac{2}{3r^2}a_1-\frac{1}{3r} \partial_r a_1\right\}\partial\beta\\
&\Longrightarrow k^{(2)}=\frac{2}{3}r^2i\omega\partial\beta.
\end{split}
\end{equation}
where the integration constant is again fixed by the condition~(\ref{frame convention}).

We are led to the large $r$ behavior for the metric correction at second order,
\begin{equation}\label{2nd order correction}
\begin{split}
k^{(2)}&=-\frac{2}{3}r^2\partial_v\partial\beta,~~~ j_i^{(2)}\xlongrightarrow{r\to\infty}-\frac{1}{3}r^2\partial_i\partial\beta+ \mathcal{O}\left(\frac{1}{r}\right),~~~\\
\alpha^{(2)}_{ij}&\xlongrightarrow{r\to\infty}\frac{2-\ln{2}}{8r^4}\partial_v\sigma_{ij}+ \mathcal{O}\left(\frac{1}{r^5} \right).
\end{split}
\end{equation}

\paragraph{Metric correction at third order.}
In order to extend previous perturbative analysis to $\mathcal{O}(\partial^3)$, we consider the following system of differential equations,
\begin{equation} \label{3rd order}
\left\{
\begin{aligned}
&0=(r^7-r^3)\partial_r^2 d_0 +(5r^6-r^2)\partial_r d_0 +2b_0-2r\partial_r b_0 -\frac{2}{3}r^3 c_0,\\
&0=r\partial_r^2 b_1-3 \partial_r b_1 +\frac{1}{3}r^3 \partial_r c_1,\\
&0=(r^7-r^3)\partial_r^2 c_2 +(5r^6-r^2)\partial_r c_2-2i\omega r^5\partial_r c_1 -r\partial_r a_2 +a_2-3i\omega r^4 c_1,\\
&0=r\partial_r^2 a_3-3 \partial_r a_3-q^2 r^3 \partial_r c_1.
\end{aligned}
\right.
\end{equation}
Since the equation for $d_0$ is of the same structure as that of $c_0$, we can solve for it in the very
same way as we did for $c_0$:
\begin{equation}\label{d0 solution}
d_0(\omega,q,r)=-\int_{r}^{\infty}\frac{dx}{x^5-x}\int_{1}^{x}\left[-\frac{2b_0(y)}{y^2} +\frac{2y\partial_yb_0(y)}{y^2}+\frac{2}{3}yc_0(y)\right]dy,
\end{equation}
where the integration constants are fixed by the boundary conditions~(\ref{AdS constraint}) and~(\ref{frame convention}).
In the large $r$ limit, $d_0$ behaves as
\begin{equation}
d_0(\omega,q,r\rightarrow \infty)=-\frac{1}{48r^4}\left(5-\pi-2\ln{2}\right) +\mathcal{O} \left(\frac{1}{r^6}\right).
\end{equation}

It is straightforward to integrate the remaining equations in~(\ref{3rd order}) over $r$ and fix the integration constants in the same way
as has been done for the lower order counterparts. For brevity, we only present the final results,
\begin{eqnarray}
\begin{split}
b_1(\omega,q,r)&=\int_r^{\infty}x^3dx \int_x^{\infty}dy\left[-\frac{1}{3y}\partial_y c_1(y)\right]\xlongrightarrow{r\to\infty}-\frac{i\omega}{15r}\left(1-\frac{\ln{2}}{2} \right),\\
c_2(\omega,q,r)&=\int_r^{\infty}\frac{dx}{x-x^5}\int_{1}^x dy \left[2i\omega y^3\partial_y c_1(y) + 3i\omega y^2 c_1(y) +\frac{\partial_y a_2(y)}{y}- \frac{a_2(y)}{y^2}\right]\\
&\xlongrightarrow{r\to\infty}\frac{1}{192r^4}\left\{6q^2+\omega^2\left[6\pi -\pi^2+ 12\left(2-3\ln{2}+\ln^2{2}\right)\right]\right\}+\mathcal{O}\left(\frac{1}{r^5}\right),\\
a_3(\omega,q,r)&=\int_r^{\infty}x^3 dx\int_x^{\infty} dy \left[\frac{q^2}{y}\partial_y c_1(y) \right]\xlongrightarrow{r\to\infty} \frac{i\omega q^2}{5r}.
\end{split}
\end{eqnarray}
We also have a third order version of eq.~(\ref{keq momentum}),
\begin{equation}
\begin{split}
&\partial_r k^{(3)}=\left\{-\frac{2}{3r^2}\left(a_2-q^2b_0\right)-\frac{1}{3r}
\left(\partial_ra_2-q^2\partial_rb_0\right)+\frac{4q^2}{9}rc_0\right\}\partial\beta\\
&\Longrightarrow k^{(3)}\rightarrow-\frac{q^2}{10r^2}\partial\beta+\mathcal{O}\left(\frac{1}{r^3}\right), ~~~\text{as}~~~r\to\infty.
\end{split}
\end{equation}

The large $r$ behavior for metric correction at order $\mathcal{O}(\partial^3)$ is,
\begin{equation}\label{3rd order correction}
\begin{split}
k^{(3)}\xlongrightarrow{r\to\infty}&\frac{1}{10r^2}\partial^2\partial\beta+\mathcal{O} \left(\frac{1}{r^3}\right),\\
j_i^{(3)}\xlongrightarrow{r\to\infty}&\frac{1}{5r} \partial_v\partial^2\beta_i +\frac{1}{15r} \left(1-\frac{1}{2} \ln{2}\right) \partial_v\partial_i\partial\beta+ \mathcal{O}\left(\frac{1}{r^2}\right),\\
\alpha_{ij}^{(3)}\xlongrightarrow{r\to\infty}&-\frac{1}{96r^4}\left\{6\partial^2+\left[  6\pi -\pi^2+ 12\left(2-3\ln{2}+\ln^2{2}\right)\right]\partial_v^2\right\}\sigma_{ij}\\
&-\frac{1}{48r^4}\left(5-\pi-2\ln{2}\right)\pi_{ij}+\mathcal{O}\left(\frac{1}{r^5}\right).
\end{split}
\end{equation}

\paragraph{Metric correction at $\mathcal{O}(\partial^n)$ with $n\geq4$.}
We end with a formal argument towards arbitrarily higher order metric correction in the derivative expansion. For $n\geq 4$, we have the following system of recursive differential equations,
\begin{equation} \label{nth order}
\left\{
\begin{aligned}
0=&(r^7-r^3)\partial_r^2d_{n-3}+(5r^6-r^4)\partial_rd_{n-3}-2i\omega r^5\partial_rd_{n-4}+\\
&\frac{q^2}{3}r^3 d_{n-5}-3i\omega r^4 d_{n-4}+2b_{n-3}-2r\partial_{r}b_{n-3}- \frac{2}{3}r^3c_{n-3},\\
0=&r\partial_r^2b_{n-2}-3\partial_rb_{n-2}+\frac{1}{3}r^3\partial_rc_{n-2}-\frac{2}{3}r^3 q^2 \partial_{r}d_{n-4},\\
0=&(r^7-r^3)\partial_r^2c_{n-1}+(5r^6-r^2)\partial_rc_{n-1}-2i\omega r^5\partial_rc_{n-2}-\\
&r\partial_ra_{n-1}+a_{n-1}-3i\omega r^4c_{n-2},\\
0=&r\partial_r^2 a_n-3 \partial_r a_n-q^2 r^3 \partial_r c_{n-2},
\end{aligned}
\right.
\end{equation}
where $d_{-1}$ should be understood as null. It can be shown by induction that large $r$ asymptotic behaviors of the coefficients $a_n$, $b_n$, $c_n$, $d_n$ ($n\geq 4$) are of universal form,
\begin{equation}
\begin{split}
&a_n(\omega,q,r)\to\frac{S_{a}^n(\omega,q)}{r}+\mathcal{O}\left(\frac{1}{r^2}\right),
~~~~~~~~~~b_{n-2}(\omega,q,r)\to\frac{S_{b}^{n-2}(\omega,q)}{r}+\mathcal{O} \left(\frac{1}{r^2}\right),\\
&c_{n-1}(\omega,q,r)\to\frac{S_{c}^{n-1}(\omega,q)}{r^4}+\mathcal{O} \left(\frac{1}{r^5}\right),~~~~
d_{n-3}(\omega,q,r)\to\frac{S_{d}^{n-3}(\omega,q)}{r^4}+\mathcal{O} \left(\frac{1}{r^5}\right).
\end{split}
\end{equation}
The $n$-th order counterpart of~(\ref{keq momentum}) is
\begin{equation}
k^{(n)}\xlongrightarrow{r\to\infty}\frac{S_{k}^n(\omega,q)}{r^2}\partial\beta+\mathcal{O} \left(\frac{1}{r^3}\right).
\end{equation}
The functions $S_a^n$ etc. are to be determined by solving the recursive equations~(\ref{nth order}), similarly as we did for the lower order metric corrections. Generically, they will take a form of fixed order polynomials in $\omega$ and $q$, $S^n=\sum_{m=0}^{m=n} \rho_m\,\omega^m\,q^{n-m}$. Although we are not able to give exact analytical expressions for $S_a^n$ etc., the formal analysis presented here is useful in obtaining the general structure of $T_{\mu\nu}$ up to arbitrary order in the derivative expansion. At any order $\mathcal{O}(\partial^n)$ with $n\geq4$, the components $j_i$ and $\alpha_{ij}$ fall off as
\begin{equation}\label{nth order correction}
j_i^{(n)}\to\frac{S_a^n(\omega,q)}{r} \beta_i+\frac{S_b^{n-2}(\omega,q)}{r} \partial_i\partial\beta,~~~
\alpha_{ij}^{(n)}\to\frac{2S_c^{n-1}(\omega,q)}{r^4}\sigma_{ij}+ \frac{S_d^{n-3}(\omega,q)}{r^4}\pi_{ij},
\end{equation}
in the large $r$ regime.

\subsection{Fluid stress tensor up to third order and beyond}\label{subsection32}
With the perturbative solutions to the metric correction at hand, we proceed by computing the fluid stress tensor~(\ref{stress tensor}). Up to the second order $\mathcal{O}(\partial^2)$,
\begin{equation}\label{stress tensor 2nd}
\left\{
\begin{aligned}
T_{00}&=3\left[1-4 b_1(x^{\alpha})\right]+\mathcal{O}(\partial^3),\\
T_{0i}&=T_{i0}=-4\beta_i(x^{\alpha})+\mathcal{O}(\partial^3),\\
T_{ij}&=\delta_{ij}\left[1-4 b_1(x^{\alpha})\right]-2 \sigma_{ij}+ (2-\ln{2})\partial_v\sigma_{ij} +\mathcal{O}(\partial^3),
\end{aligned}
\right.
\end{equation}
which is exactly the results obtained in~\cite{0712.2456} when linearized as in~(\ref{linearization}). Let us  write the fluid stress tensor as a formal derivative expansion,
\begin{equation}
T_{\mu\nu}=\sum_{n=0}^{\infty}T_{\mu\nu}^{(n)},
\end{equation}
where the zeroth order $T_{\mu\nu}^{(0)}$ corresponds to the non-derivative terms in eq.~(\ref{stress tensor 2nd}), which can be uplifted to the standard form of~(\ref{ideal stress tensor}). At third order, nonzero components of the fluid stress tensor are
\begin{equation}
\begin{split}
T_{ij}^{(3)}=&\frac{1}{24}\left\{6\partial^2+\left[6\pi -\pi^2+ 12\left(2-3\ln{2}+ \ln^2{2}\right)\right]\partial_v^2\right\}\sigma_{ij}\\
&+\frac{1}{12}\left(5-\pi-2\ln{2}\right)\pi_{ij},
\end{split}
\end{equation}
where the tensor structure $\pi_{ij}$ appears for the first time. From the expression for the stress tensor we can immediately read off the viscosity functions $\eta$ and $\zeta$ as quoted in (\ref{viscosity funs}).

Substituting eq.~(\ref{nth order correction}) into eq.~(\ref{stress tensor}), we arrive at a general form of $T_{\mu\nu}$ at arbitrary order in the gradient expansion,
\begin{equation}\label{stress tensor higher orders}
\begin{split}
\tilde{T}_{ij}^{(n)}&\to\frac{1}{2r^2}\times(-2r)(r^6-r^2)\partial_r\alpha_{ij}^{(n)} \\ \Longrightarrow T_{ij}^{(n)} &= 8S_c^{n-1}(\omega,q)\cdot\sigma_{ij}+ 4S_d^{n-3}(\omega,q)\cdot\pi_{ij}.
\end{split}
\end{equation}
This gives formal expressions for the viscosity functions as sums over the coefficients $S^n$: $\eta= -\sum_n 8S_c^{n-1}$ and $\zeta=- \sum_n 4S_d^{n-3}$.
\section{All order linearized hydrodynamics}\label{section4}
To fully account for all order derivative terms in the fluid stress tensor, we resort to
numerical techniques for solution of the RG equations~(\ref{abcd eqs}), extending validity of the above discussed hydrodynamic regime to large momenta. It is convenient to rescale the functions $a(r)$ and $b(r)$
\begin{equation}\label{renormalization}
a(\omega,q,r)= r^4\tilde{a}(\omega,q,r),~~~~b(\omega,q,r)= r^4\tilde{b}(\omega,q,r),
\end{equation}
and also use $u$-coordinate instead of $r$
\begin{equation}\label{u coordinate}
u\equiv \frac{1}{r}\Longrightarrow u\in[0,1].
\end{equation}
In $u$ coordinate, the horizon is located at $u=1$ while the conformal boundary is at $u=0$. In what follows, we also use notations $\tilde{c}(u)=c(r)$ and $\tilde{d}(u)=d(r)$ to stress that they are functions of $u$. Equations~(\ref{abcd eqs}) become
\begin{equation}\label{abcd eqs new}
\left\{
\begin{aligned}
&0=u\tilde{a}^{\prime\prime}-3\tilde{a}^{\prime} +q^2u\tilde{c}^{\prime}-3i\omega- q^2u,\\
&0=u\tilde{b}^{\prime\prime}-3\tilde{b}^{\prime}-\frac{1}{3}u\tilde{c}^{\prime}+ \frac{2}{3}q^2 u \tilde{d}^{\prime}-u,\\
&0=(u-u^5)\tilde{c}^{\prime\prime}-(3+u^4-2 i\omega u)\tilde{c}^{\prime}+ (u\tilde{a}^{\prime}-3\tilde{a})-3 i\omega \tilde{c} +3-i\omega u,\\
&0=(u-u^5)\tilde{d}^{\prime\prime}-(3+u^4-2 i\omega u)\tilde{d}^{\prime}+ 2(u\tilde{b}^{\prime}-3\tilde{b})+ \left(\frac{1}{3} q^2u-3i\omega \right)\tilde{d} -\frac{2}{3}u \tilde{c},
\end{aligned}
\right.
\end{equation}
where prime denotes derivative with respect to $u$. The problem of resumming all
order derivative terms in the boundary stress tensor is reduced to a boundary value problem of the system of ordinary differential equations~(\ref{abcd eqs new}).

In the rest of this chapter we will solve this problem by two methods. The first method will be fully numerical while the second one is an approximate analytic scheme. Both methods are demonstrated to converge to the same results.

\subsection{Numerical results from the shooting method}\label{subsection41}
We have to impose boundary conditions both at the horizon and asymptotic infinity. We apply a \emph{shooting method} to solve the system~(\ref{abcd eqs new}). The main idea behind the shooting method is to reduce the boundary value problem to an initial value problem for a system such as~(\ref{abcd eqs new}). One starts from a trial solution (initial condition) at one boundary (horizon) and integrates the system until the other boundary. Then, thus obtained solution should be matched with boundary conditions at the end of the integration. That would not happen for an arbitrary trial initial condition: the trial solution has to be fine tuned in order for the boundary conditions at the end of the integration to be satisfied. This fine tuning problem can be turned into an optimization procedure.

We are now to discuss an implementation of this method for the system~(\ref{abcd eqs new}) given the boundary conditions presented in section~\ref{section2}. In order to fully find a solution for four second order differential equations, we have to specify overall eight boundary conditions. The regularity requirement at horizon, boundary conditions~(\ref{AdS constraint}) and the Landau frame convention~(\ref{frame convention}), indeed do provide precisely eight conditions: two at the horizon and six at the conformal boundary.

\paragraph{Series solution near the horizon.}
We start from the regularity requirement at the unperturbed horizon $u=1$. To have a regular black hole solution  near $u=1$, the functions $\tilde{a},\tilde{b},\tilde{c},\tilde{d}$ have to be Taylor expandable,
\begin{equation}\label{horizon data}
\begin{split}
\tilde{a}(\omega,q,u)=\sum_{n=0}^{\infty}A_{h}^n(u-1)^n,~~~~\tilde{b} (\omega,q,u)= \sum_{n=0}^{\infty} B_{h}^n(u-1)^i,\\
\tilde{c}(\omega,q,u)=\sum_{n=0}^{\infty}C_{h}^n(u-1)^n,~~~~\tilde{d} (\omega,q,u)= \sum_{n=0}^{\infty} D_{h}^n(u-1)^n,
\end{split}
\end{equation}
where the subscript ``$h$'' indicates that eq.~(\ref{horizon data}) is a series solution near \emph{horizon}. The regularity condition at $u=1$ fixes only two integration constants in these four functions. This is consistent with the observation that $u=1$ singular point in the equations for $a(u)$ and $b(u)$ is due to $c(u)$ and $d(u)$.
Six coefficients $A_{h}^0$, $A_h^1$, $B_h^0$, $B_h^1$, $C_h^0$ and $D_h^0$ remain unconstrained. The rest of the coefficients in~(\ref{horizon data}) can be
expressed in terms of these six coefficients via substitution of the series~(\ref{horizon data}) into the system~(\ref{abcd eqs new}).

\paragraph{Series solution near the conformal boundary.}
We turn to discuss near $u=0$ behavior for these functions. At $u=0$, the characteristic indices for the system~(\ref{abcd eqs new}) are $0$ and $4$. Series solution then takes the form,
\begin{equation}\label{boundary data}
\begin{split}
\tilde{a}(\omega,q,u)=\sum_{n=0}^{\infty}A_b^n u^n +u^4\ln{u}\sum_{n=0}^{\infty}A_L^n u^n,~
\tilde{b}(\omega,q,u)=\sum_{n=0}^{\infty}B_b^n u^n +u^4\ln{u}\sum_{n=0}^{\infty}B_L^n u^n,\\
\tilde{c}(\omega,q,u)=\sum_{n=0}^{\infty}C_b^n u^n +u^4\ln{u}\sum_{n=0}^{\infty}C_L^n u^n,~
\tilde{d}(\omega,q,u)=\sum_{n=0}^{\infty}D_b^n u^n +u^4\ln{u}\sum_{n=0}^{\infty}D_L^n u^n,
\end{split}
\end{equation}
where the subscript ``$b$'' marks the asymptotic infinity $u=0$. The logarithmic branch, whose coefficients are labeled with the subscript ``$L$'', is necessary due to the fact that the difference between two characteristic indices is \emph{integer}. Similarly to the near horizon expansion, by substituting~(\ref{boundary data}) into~(\ref{abcd eqs new}), all the coefficients of~(\ref{boundary data}) are related to the following eight coefficients $A_{b}^0$, $A_b^4$, $B_b^0$, $B_b^4$, $C_b^0$, $C_b^4$, $D_b^0$ and $D_b^4$.

The boundary condition~(\ref{AdS constraint}) implies that
\begin{equation}\label{AdS constraint1}
\begin{split}
A_b^0=B_b^0=C_b^0=D_b^0=0.
\end{split}
\end{equation}
while the ``Landau frame'' convention~(\ref{frame convention}) constrains two more  expansion coefficients
\begin{equation}\label{frame convention1}
A_b^4=B_b^4=0.
\end{equation}
leaving only two undetermined coefficients, $C_b^4$ and $D_b^4$, which have to be determined through dynamical evolution from the horizon.

With the conditions~(\ref{AdS constraint1}) and~(\ref{frame convention1}) at hand, the logarithmic branch in~(\ref{boundary data}) vanishes identically and the large $r$ behavior for these functions is thus
\begin{equation}
\left\{
\begin{aligned}
\tilde{a}(\omega,q,u)&=-i\omega u+\mathcal{O}\left(u^5\right),~~~~~~~~ \tilde{b}(\omega,q,u)=-\frac{1}{3} u^2+ \mathcal{O} \left(u^5\right),\\
\tilde{c}(\omega,q,u)&=u+C_b^4 u^4+\mathcal{O}\left(u^5\right),~~~
\tilde{d}(\omega,q,u)=D_b^4 u^4+\mathcal{O} \left(u^5\right).
\end{aligned}
\right.
\end{equation}
In terms of functions $a,~b,~c,~d$, we have
\begin{equation}
\left\{
\begin{aligned}
a(\omega,q,r)&=-i\omega r^3+\mathcal{O}\left(\frac{1}{r}\right),~~~~~~~
b(\omega,q,r)=-\frac{1}{3}r^2+\mathcal{O} \left(\frac{1}{r}\right),\\
c(\omega,q,r)&=\frac{1}{r}+\frac{C_b^4}{r^4}+\mathcal{O}\left(\frac{1}{r^5}\right),~~~
d(\omega,q,r)= \frac{D_b^4}{r^4}+\mathcal{O}\left(\frac{1}{r^5}\right).
\end{aligned}
\right.
\end{equation}
The coefficient functions $C_b^4$ and $D_b^4$ can be now identified with the viscosity functions $\eta(\omega,q^2)$ and $\zeta(\omega,q^2)$
\begin{equation}
\eta(\omega,q^2)=-8C_b^4,~~~~~~~~~\zeta(\omega,q^2)=-4D_b^4.
\end{equation}
Our problem is now mapped into finding such six near horizon expansion coefficients $A_h^0$,~$A_h^1$,~$B_h^0$,~$B_h^1$,~$C_h^0$ and $D_h^0$ that would make six boundary coefficients $A_b^0$,~$A_b^4$,~$B_b^0$,~$B_b^4$,~$C_b^0$, $D_b^0$ to vanish in accordance with the above boundary conditions. Once this is achieved, the coefficients $C_b^4$ and $D_b^4$ should be read off from the final solution. Therefore, the boundary value problem for the system~(\ref{abcd eqs new}) is reduced to the problem of \emph{root-finding} or \emph{optimization} in numerical analysis,
\begin{equation}\label{root finding}
\left\{A_b^0,~A_b^4,~B_b^0,~B_b^4,~C_b^0,~D_b^0\right\}[A_h^0,A_h^1,B_h^0,B_h^1,C_h^0,D_h^0]=0.
\end{equation}
The procedure is repeated for each value of  $\omega$ and $q$.

\paragraph{Further numerical details and results.}

The near-horizon series solution~(\ref{horizon data}) makes it possible to evaluate
\begin{equation}
\left\{\tilde{a},~\tilde{a}^{\prime},~\tilde{b},~\tilde{b}^{\prime},~\tilde{c},~
\tilde{c}^{\prime},~\tilde{d},~\tilde{d}^{\prime}\right\}(\omega,q,u)
\end{equation}
at some point $u=1-u_{\star}$, close to the  horizon ($u_{\star}\ll 1)$. That helps to avoid a numerically
problematic region near horizon where the system of equations has a singular point.
In our shooting routine, we integrate the system~(\ref{abcd eqs new}) from this
near-horizon point $u=1-u_{\star}$ till some point $u=u_{\star}$, close to asymptotic infinity $u=0$.

We use Newton's method~\cite{numbook} for root-finding  and find it works rather well.
Efficient initial guesses for the trial values of $A_h^0$,~$A_h^1$,~$B_h^0$,~$B_h^1$,~$C_h^0$ and $D_h^0$
are provided by linearly extrapolating previously  computed roots along the $\omega$ or $q^2$-axis.
The numerical procedure is started  from the point $\omega=q^2=0$, known exactly from  section~\ref{section3}
\begin{equation}
\begin{split}
A_h^0&=A_h^1=0,\\
B_h^0&=-\frac{3}{8},~~B_h^1=-\frac{1}{24}\left(16+\pi+2\ln{2}\right)\approx -0.85533\cdots,\\
C_h^0&=\frac{1}{8}\left(\pi+6\ln{2}\right)\approx 0.91256\cdots,\\
D_h^0&=\frac{1}{576}\left(48\mathcal{C}-18\pi+\pi^2+108\ln{2}-48\ln^2{2}-24 \pi \ln{2}\right)\approx -0.0055149\cdots,
\end{split}
\end{equation}
where $\mathcal{C}$ is the Catalan constant with approximate value $\mathcal{C}\approx 0.915966$. We set $u_{\star}$ to $10^{-5}$ and checked stability of the results with respect to variations of $u_{\star}$.

\begin{figure}[h]
\includegraphics[scale=0.58]{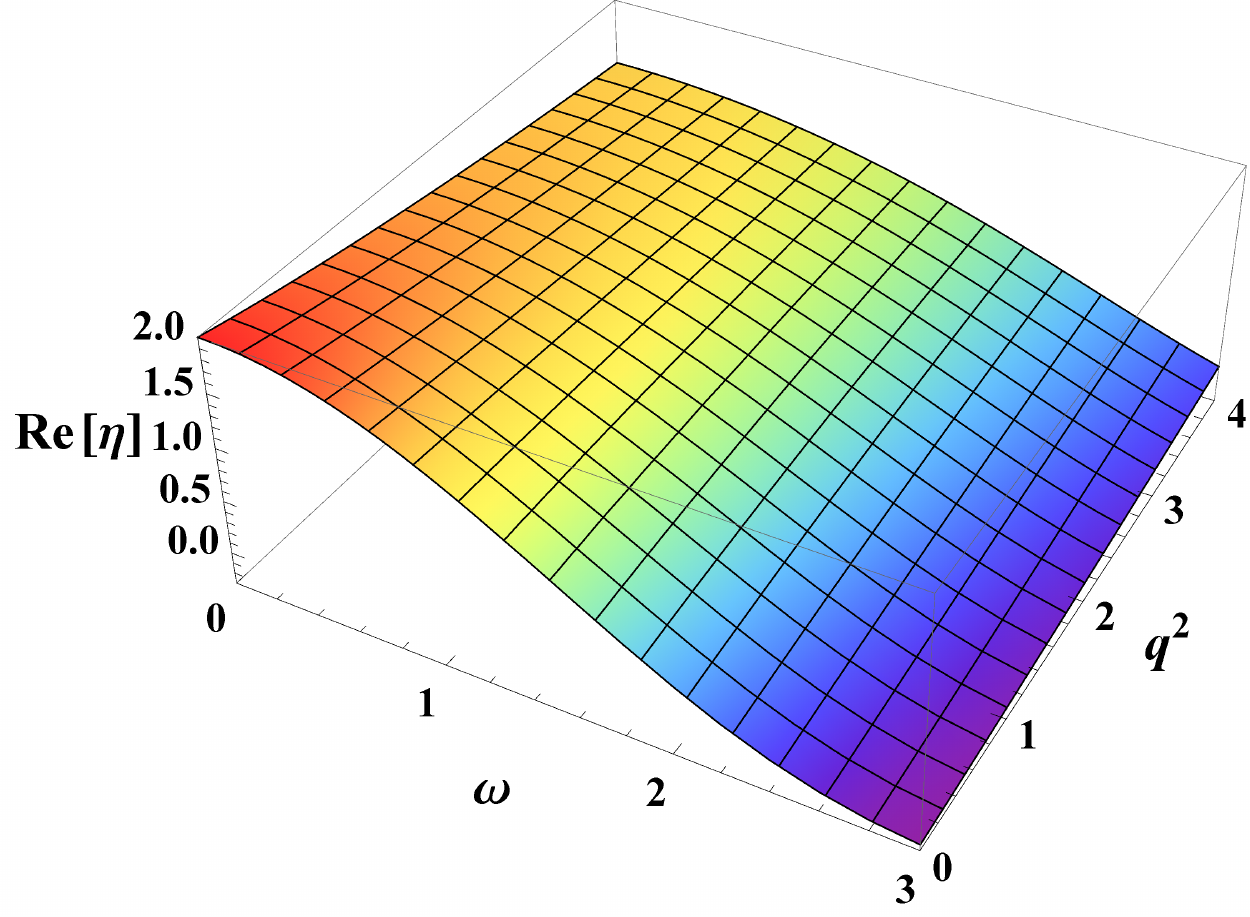}
\includegraphics[scale=0.58]{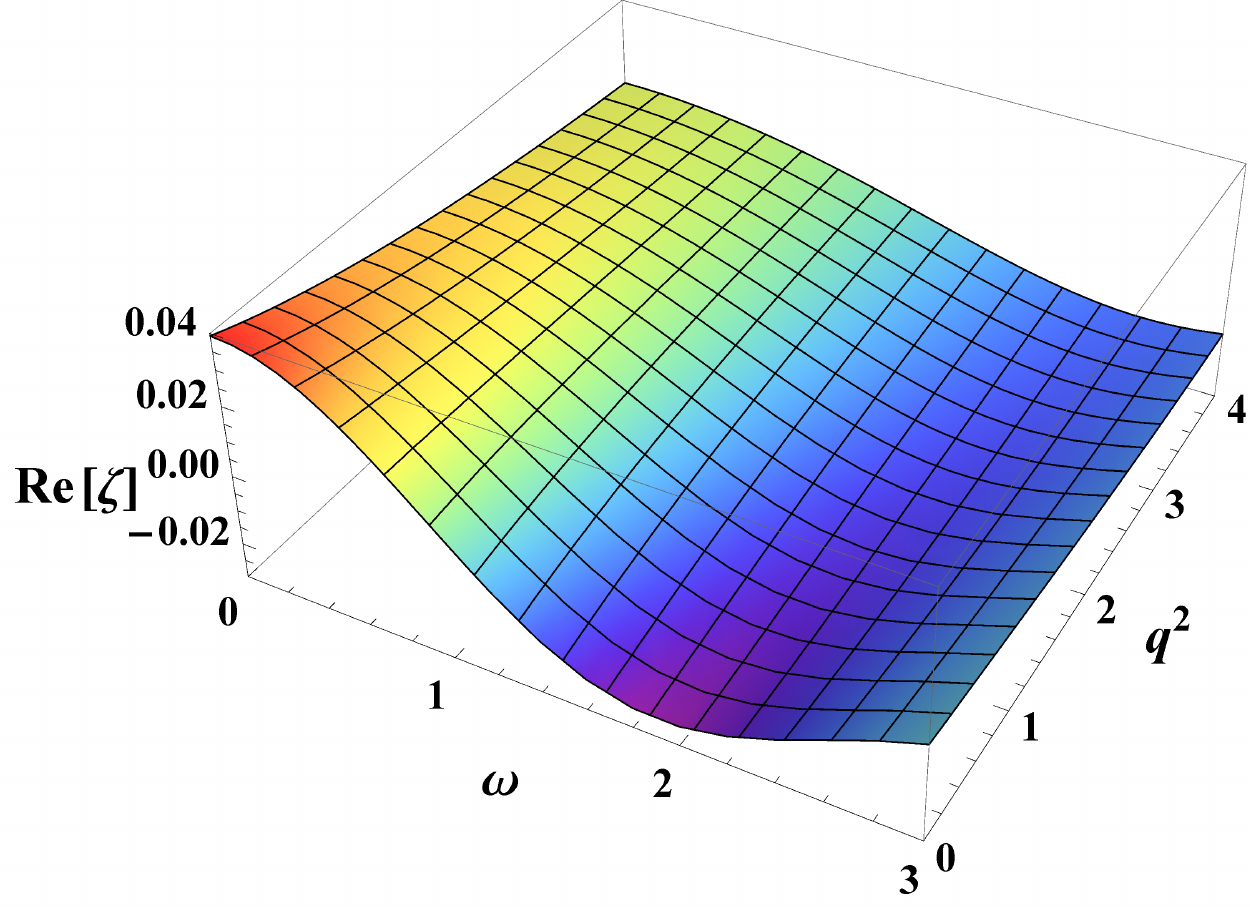}
\includegraphics[scale=0.58]{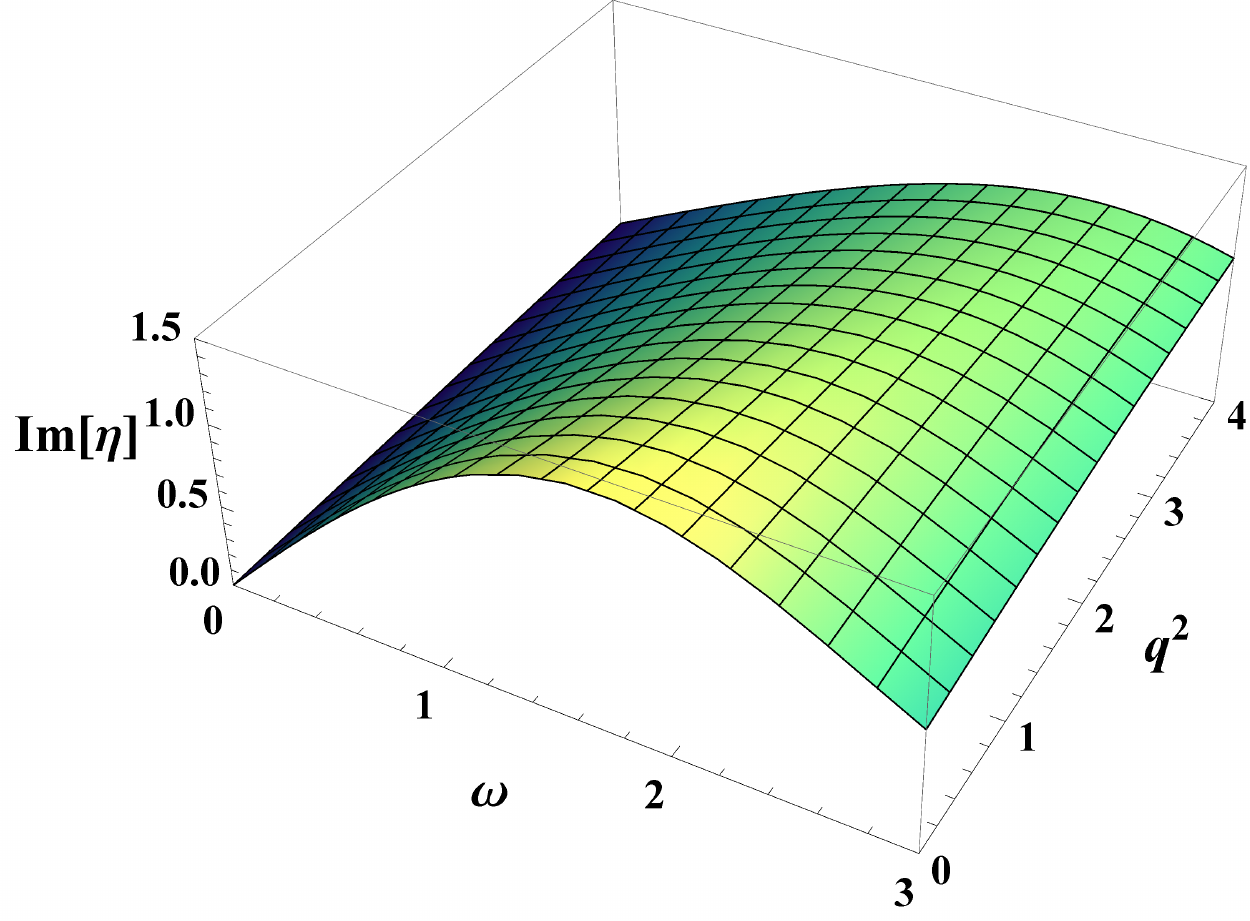}
\includegraphics[scale=0.58]{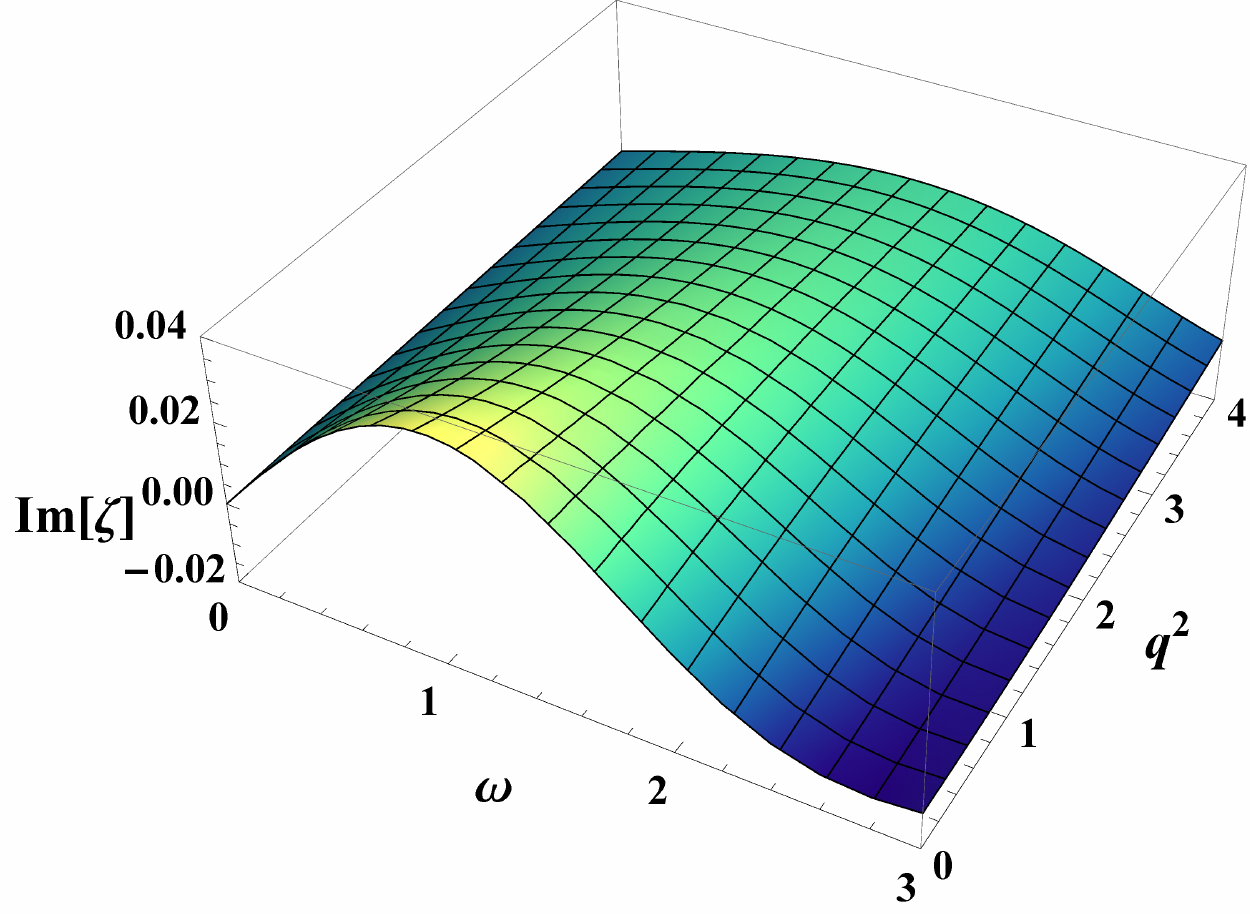}
\caption{The viscosities $\eta(\omega,q^2)$ and $\zeta(\omega,q^2)$ as functions of $\omega$ and $q^2$.}
\label{figure1}
\end{figure}

Our numerical results for the viscosity functions $\eta(\omega,q^2)$ and $\zeta(\omega,q^2)$ are shown in Figure~\ref{figure1}.
The real parts of $\eta$ and $\zeta$ decrease with momenta until reaching  minima around points $\left\{\omega\approx 3.0,q^2=0\right\}$ and $\left\{\omega\approx 1.9,q^2=0\right\}$, respectively. A sign of damped oscillations is observed in the results, while eventually, the real parts vanish  at very
large $\omega$ and/or $q^2$. The imaginary parts of the viscosities first increase from zero up to some maxima around  $\left\{\omega\approx 1.7,q^2=0\right\}$ for $\eta$ and $\left\{\omega\approx 1.0,q^2=0\right\}$ for $\zeta$. With further increase of the momenta, the imaginary parts decrease  reaching  zero at  large momenta.

\begin{figure}[h]
\includegraphics[scale=0.86]{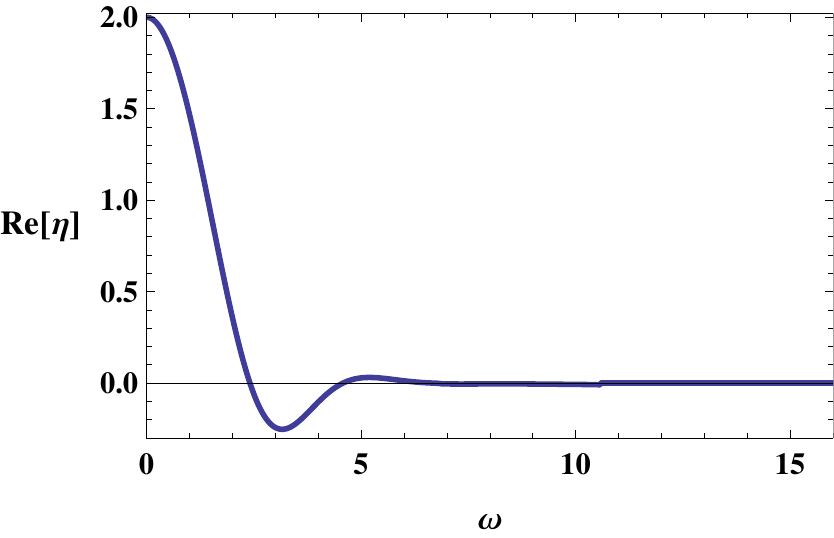}
\includegraphics[scale=0.88]{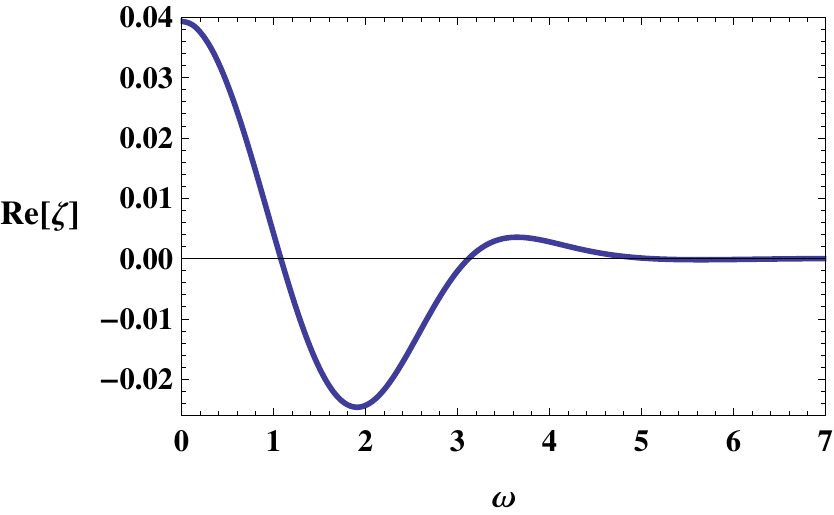}
\includegraphics[scale=0.86]{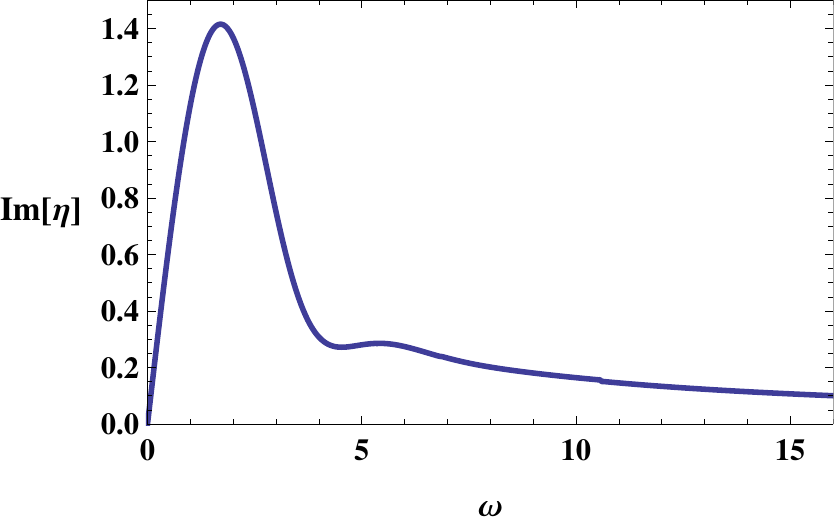}
\includegraphics[scale=0.88]{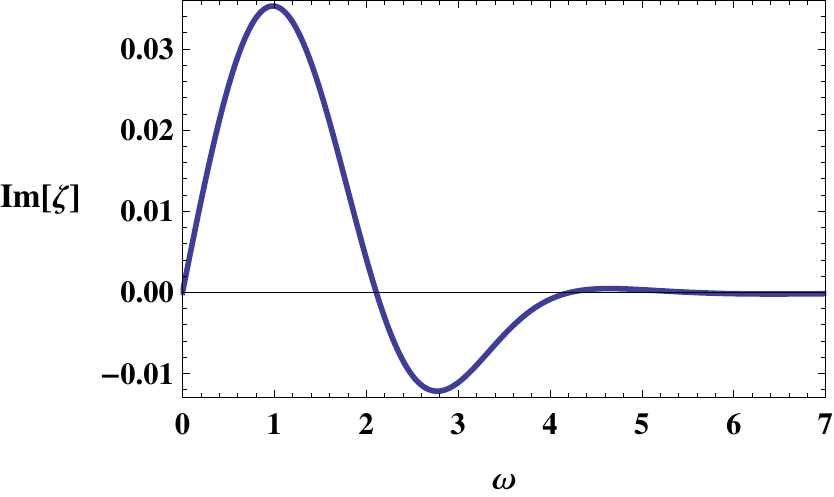}
\caption{The generalized viscosity functions $\eta$ and $\zeta$ vs $\omega$ at $q^2=0$.}
\label{figure2}
\end{figure}

\begin{figure}[h]
\includegraphics[scale=0.88]{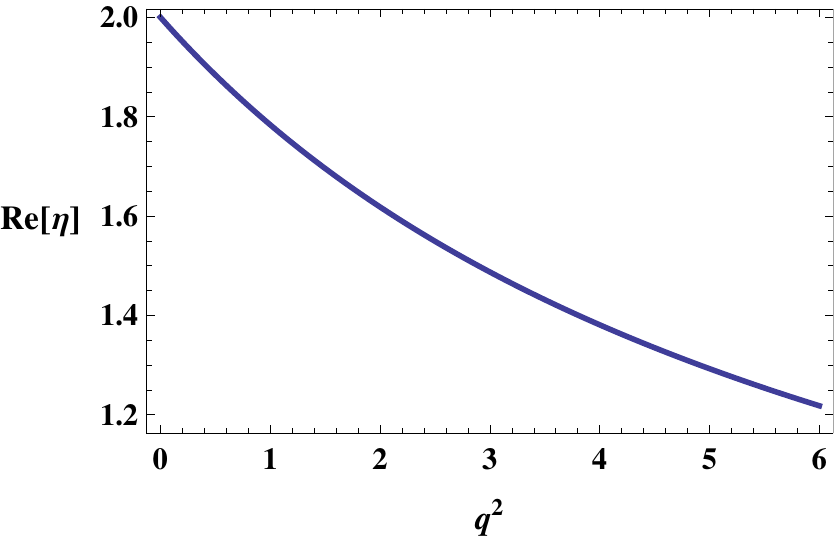}
\includegraphics[scale=0.88]{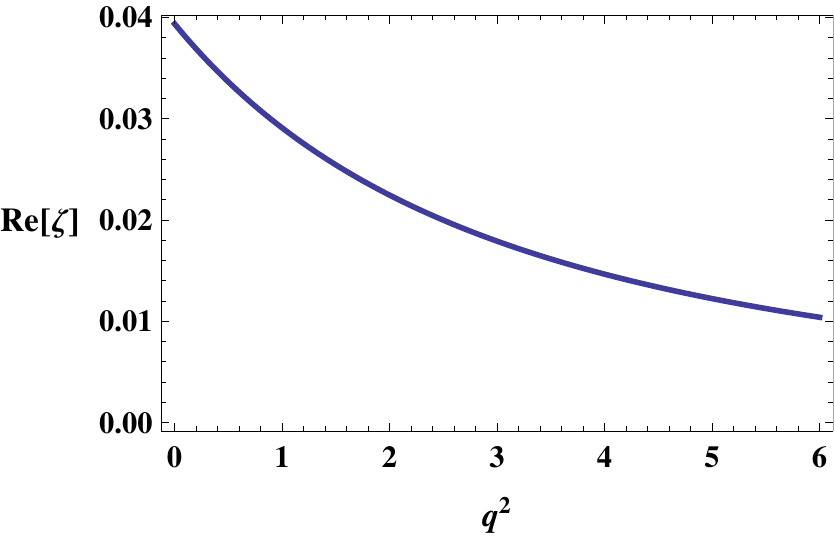}
\caption{The generalized viscosity functions $\eta$ and $\zeta$ vs  $q^2$ at $\omega=0$.}
\label{figure3}
\end{figure}

Vanishing of transport coefficients at very large momenta is well anticipated:  there should be no response at very short times or distances. This point is critical for the generalized relativistic hydrodynamics to be causal. To further confirm our observation, we focus on large momenta behavior for the viscosity functions. In Figures~\ref{figure2} and~\ref{figure3}, we show our results for very large $\omega$ or $q^2$.
The imaginary parts of $\eta$ and $\zeta$ are identically zero at $\omega=0$. This is obvious from the eqs.~(\ref{abcd eqs new}), which have no imaginary terms at $\omega=0$. Vanishing of the viscosity functions at large $\omega$ (and/or $q^2$) is an important factor for reliably addressing early time stage in heavy ion collisions~\cite{1302.0697,0704.1647,1108.3972}.

The imaginary part of $\eta$ never turns negative. This is indeed a necessary condition for stability of the fluid equations, as is seen from the shear mode dispersion relation~(\ref{shear mode}). In contrast, the imaginary part of $\zeta$ does change sign at intermediate values of $\omega$ and $q^2$ as seen in Figure~\ref{figure1}. However, negative $\zeta$ does not mean an instability: it only contributes as a practically negligible correction to the dispersion relation of sound mode~(\ref{sound mode}), but never turns any of these modes into unstable. Indeed, the absolute value of $\zeta(\omega,q^2)$ is always highly suppressed as compared to that of $\eta(\omega,q^2)$. Furthermore, the inequality $q^2\zeta(\omega,q^2)\ll \eta(\omega,q^2)$ is valid in all the kinematic range covered by our numerical results. Therefore, for any practical applications and hydrodynamic modelings, it is probably always safe to ignore the viscosity function $\zeta(\omega,q^2)$.

\subsection{Approximate results from the matching method}\label{subsection42}
In this subsection we provide an alternative approach to solving equations~(\ref{abcd eqs new}) based on an analytic approximate scheme. This provides us with a possibility to check the numerical results of the previous subsection. The main idea is to adopt a matching method, which was introduced in~\cite{0907.3203} in order to provide analytical evidence for condensation phenomena in a holographic superconductor model~\cite{0803.3295}. The method is based on the series expansions ~(\ref{horizon data}) and~(\ref{boundary data}), which not only exactly solve the system~(\ref{abcd eqs new}) but also should match over the whole regime of $u\in [0,1]$.

\begin{figure}[h]
\includegraphics[scale=0.58]{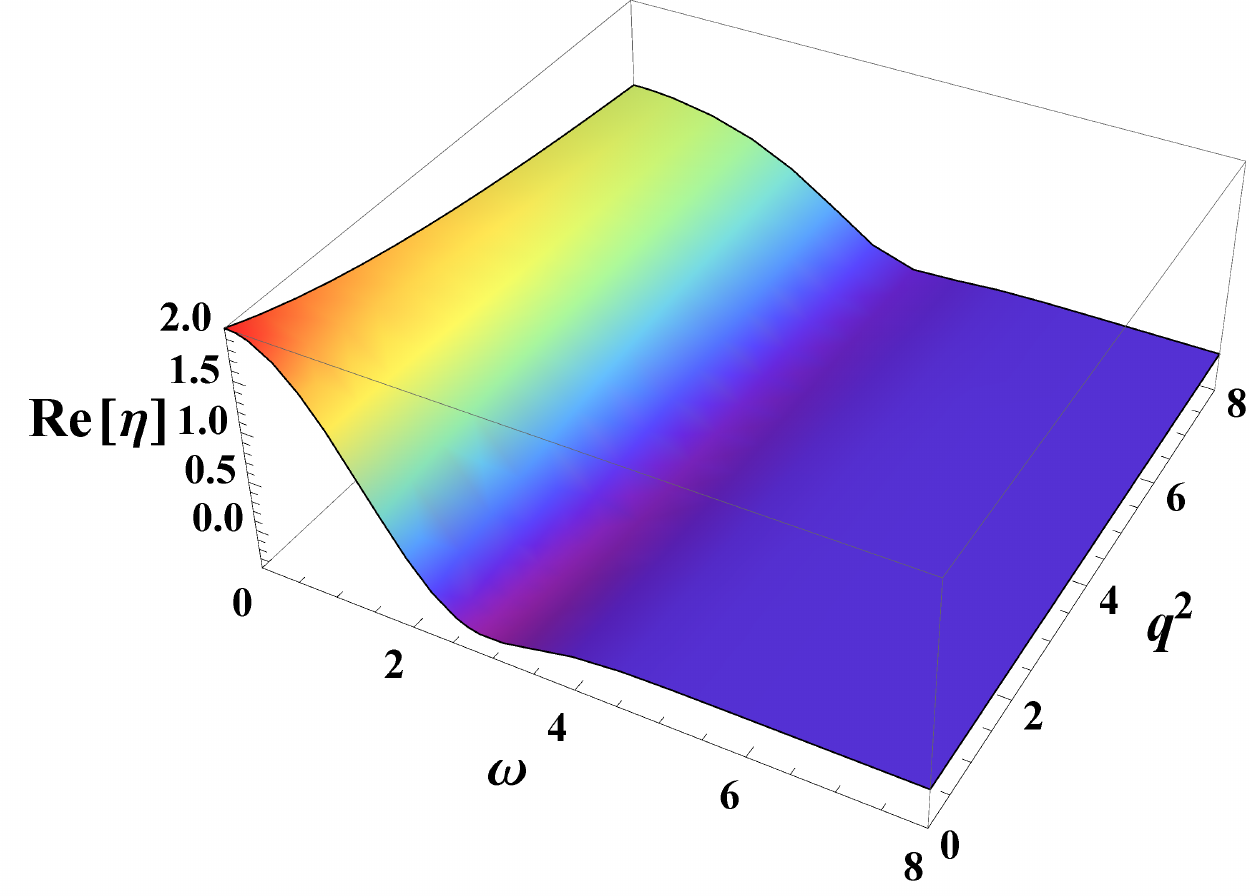}
\includegraphics[scale=0.58]{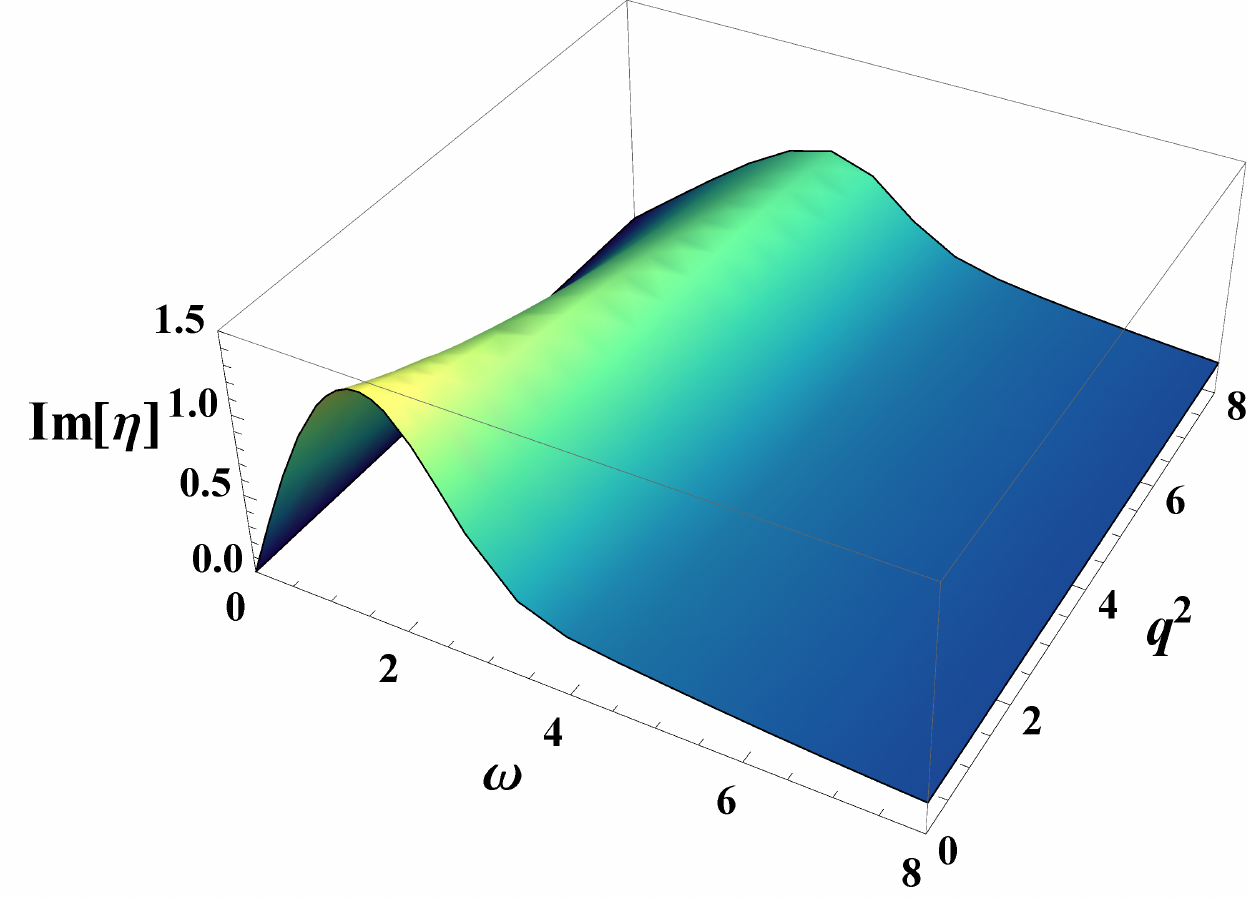}
\includegraphics[scale=0.58]{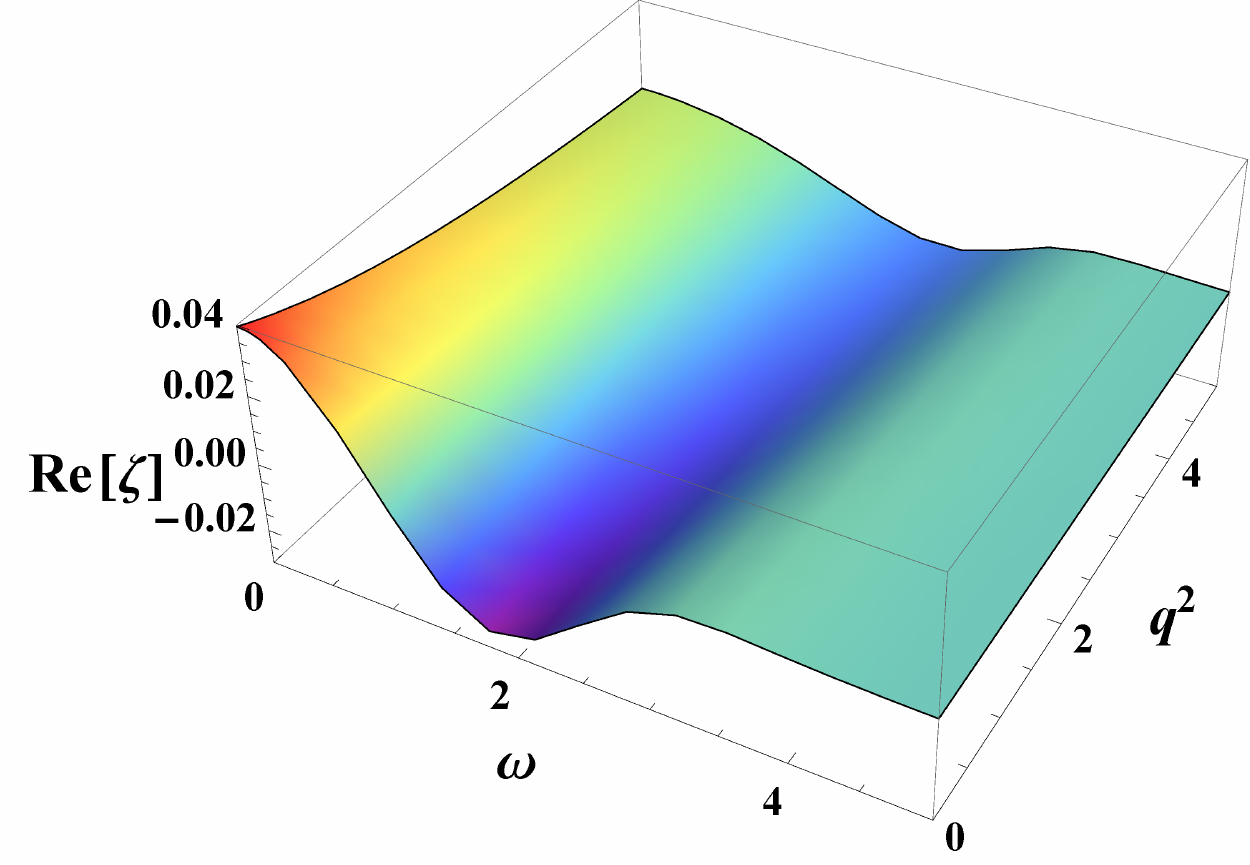}
\includegraphics[scale=0.58]{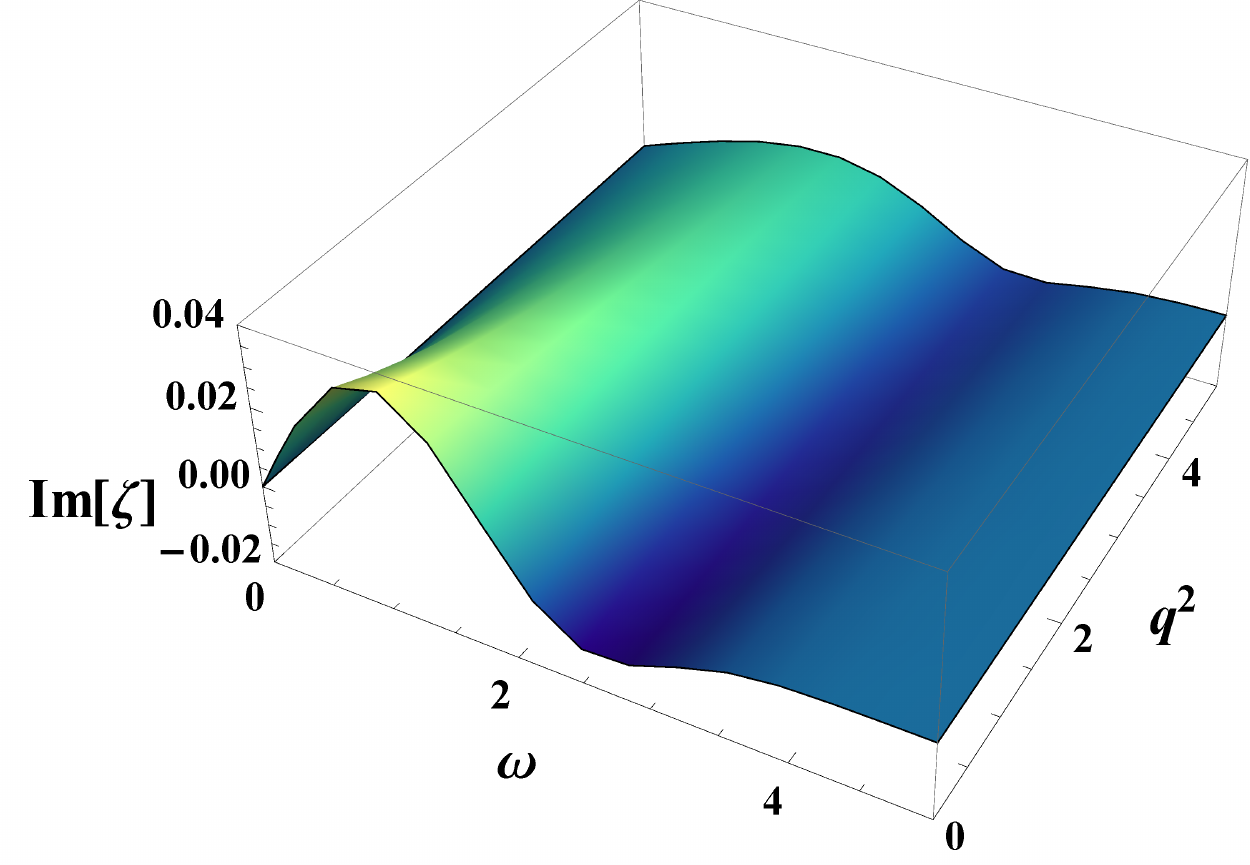}
\caption{The viscosity functions $\eta(\omega,q^2)$ and $\zeta(\omega,q^2)$ vs $\omega$ and $q^2$ from the matching method.}
\label{figure4}
\end{figure}

The approximation we are to employ is a truncation of the series~(\ref{horizon data}) and~(\ref{boundary data}): we will keep up to eleven terms in each expansion, i.e., order $u^{10}$ and $(u-1)^{10}$, respectively. While the truncated series would not any more solve the system~(\ref{abcd eqs new}) exactly, keeping enough terms guarantees accurate solutions near the horizon and conformal boundary. We then match the truncated series solutions at an intermediate point such as $u=1/2$. Taking $\tilde{a}(\omega,q,u)$ as an example, the matching of its value and first order derivative at $u=1/2$ results in,
\begin{equation}\label{matchinga}
\begin{split}
\sum_{n=0}^{10} A_h^n (u-1)^n\Bigg|_{u=1/2}&=\left(\sum_{n=0}^{10} A_b^n u^n+u^4 \ln{u}\sum_{n=0}^{10} A_L^n u^n\right)\Bigg|_{u=1/2},\\
\left(\sum_{n=0}^{10} A_h^n (u-1)^n\right)^{\prime}\Bigg|_{u=1/2}&=\left(\sum_{n=0}^{10} A_b^n u^n+u^4 \ln{u}\sum_{n=0}^{10} A_L^n u^n\right)^{\prime}\Bigg|_{u=1/2},
\end{split}
\end{equation}
where the boundary conditions~(\ref{AdS constraint1}) and~(\ref{frame convention1}) should be imposed. This method casts the system~(\ref{abcd eqs new}) into algebraic equations for these expansion coefficients. On the one hand, having kept a large number (eleven) of terms in the expansions, we achieve stable and highly accurate results. The viscosity functions obtained from the matching scheme are displayed in Figure~\ref{figure4}.

On the other hand, the large number of terms kept in~(\ref{matchinga}) lead to analytical but rather lengthy and not very illuminating expressions for the viscosity functions. They are of the type
\begin{equation}\label{hydro limit}
\eta(\omega,q^2)\approx\frac{P_{\eta}(\omega,q^2)}{Q_{\eta}(\omega,q^2)},~~~~~~~~~
\zeta(\omega,q^2) \approx\frac{P_{\zeta}(\omega,q^2)}{Q_{\zeta}(\omega,q^2)}.
\end{equation}
where $P$s and $Q$s are finite order polynomials in $\omega$ and $q^2$. Below, we only quote a few terms in hydrodynamic expansion of
these polynomials,
\begin{equation}
\begin{split}
P_{\eta}(\omega,q^2)=&2.00-1.82 i\omega - 1.01\omega^2 + 0.249 q^2-0.208 i\omega q^2+ 0.401 i \omega^3-0.103\omega^2 q^2\\
& + 0.124 \omega^4+ 0.0119 q^4+\cdots,\\
Q_{\eta}(\omega,q^2)=&1.00-1.56 i\omega- 1.24\omega^2 + 0.250 q^2 -0.296 i \omega q^2
+0.656 i\omega^3-0.189\omega^2 q^2\\
& + 0.255\omega^4+ 0.0178q^4++0.0822 i\omega^3 q^2 -0.0780 i \omega^5 +\cdots,\\
P_{\zeta}(\omega,q^2)=&0.0396 -0.0666 i\omega -0.0611\omega^2 +0.00772q^2 +\cdots,\\
Q_{\zeta}(\omega,q^2)=&1.00-3.13i\omega-4.93\omega^2+ 0.526 q^2- 1.44 i\omega q^2
+ 5.20i \omega^3-2.01 \omega^2 q^2\\
&+ 4.11 \omega^4+0.107 q^4 +1.89i \omega^3 q^2-2.59i \omega^5+\cdots.
\end{split}
\end{equation}
The structure~(\ref{hydro limit}) implies that the viscosities have a number of poles (zeros of $Q$s) and this is quite consistent with the arguments made in~\cite{0905.4069} that exact $\eta(\omega,q^2)$ in fact has infinitely many poles.

In the hydrodynamic limit, our approximate results for the viscosities are
\begin{equation}\label{visexp}
\begin{split}
\eta_{\text{m}}(\omega,q^2)&=2.00+ 1.31i\omega-0.567\omega^2-0.252 q^2-0.337i\omega q^2 \\
&-0.169i\omega^3+0.243\omega^2 q^2+ 0.0306\omega^4 + 0.0395 q^4+\cdots,\\
\zeta_{\text{m}}(\omega,q^2)&=0.0396 + 0.0573i\omega - 0.0449 \omega^2- 0.0131 q^2 +\cdots,
\end{split}
\end{equation}
They are in perfect agreement with the analytical results of section~\ref{section3} as quoted in (\ref{viscosity funs}), with up to $1\%$ error.
We notice that the expansion (\ref{visexp}) provides an extension of our analytical third order result to much higher order.

The obtained approximate results make it possible to explore the asymptotic behavior of the viscosity functions in the limit of very large momenta,
\begin{equation}\label{asym beh}
\eta(\omega,q^2)\sim \frac{i}{\omega},~~~~\zeta(\omega,q^2)\sim -\frac{i}{\omega^3}, ~~~\text{as}~~~\omega\to \infty~\text{and}~q^2\to \infty.
\end{equation}
This asymptotic behavior of~(\ref{asym beh}) provides another confirmation for vanishing of $\eta$ and $\zeta$ in the large momenta regime.

\section{Summary and discussion}\label{section5}
In this paper we have provided all the details and expanded presentation of the results  advertised in our short publication~\cite{1406.7222}. As a further development of the ideas put forward in~\cite{0704.1647}, we have consistently determined the energy-momentum stress tensor of a weakly perturbed conformal fluid, whose underlying microscopic description is a strongly coupled $\mathcal{N}=4$ super-Yang-Mills theory at finite temperature. The results were derived by linearizing the fluid/gravity correspondence. We have included all order derivative terms in the boundary stress tensor, limiting the study to small amplitude perturbations only. We have found that all order dissipative terms in the fluid stress tensor are fully accounted for by two (generalized) momenta-dependent viscosity functions $\eta(\omega,q^2)$ and $\zeta(\omega,q^2)$. $\eta(\omega,q^2)$ is a transport coefficient in front of the shear tensor $\sigma_{ij}$ while $\zeta(\omega,q^2)$ is a coefficient of a new tensor $\pi_{ij}$, which is given in terms of third order derivative of the fluid velocity.

As one of our main results, we have derived a closed-form \emph{linear} holographic RG flow-type equations~(\ref{abcd eqs}) for the viscosity functions. Intriguingly, an analogous holographic RG flow equation for electrical conductivity obtained in~\cite{0809.3808} is a \emph{nonlinear} one. The constraint components of the bulk Einstein equations~(\ref{einstein eq}) have been shown to generalize the Navier-Stokes equations, consistently with the conservation laws of the fluid stress tensor. We have analytically computed the viscosity functions, up to third order in the hydrodynamic gradient expansion, and the dispersion relations for the shear and sound waves. These third order corrections are needed in order to correctly reproduce the dispersion relation for the shear wave up to order $\mathcal{O}(q^4)$, as emphasized in~\cite{0712.2451,0712.2456}.

To include all order dissipative effects in the fluid stress tensor, we have  solved numerically the RG flow-type equations~(\ref{abcd eqs}). The numerical results for the viscosity functions are displayed in Figure~\ref{figure1}.  Based on our numerical calculations,   we have been able to significantly extend
knowledge about the viscosity functions in the hydrodynamic limit, providing  in (\ref{visexp}) an expansion
up to fifth order.
Consistently with physical intuition, the viscosities  vanish at very large momenta as seen in Figures~\ref{figure2} and~\ref{figure3}. We have verified our results by solving the equations~(\ref{abcd eqs}) using an alternative method.

Importantly, the hydrodynamic theory constructed in this work is causal and should be free of any instabilities if implemented as a dynamical model for plasma evolution.
The viscosity function encodes an infinite set of quasi-normal modes \cite{hep-th/0506184} including corresponding residues.

Obviously, this is not QCD, but we hope that some generic features about momenta-dependent transport coefficients and high gradient structures have been captured by our results. They can help in building new models of causal relativistic hydrodynamics, beyond Navier-Stokes or Israel-Stewart.

The stress tensor computed in this work resums all order gradients linearized in the amplitude of fluid velocity and is not readily applicable to phenomena where nonlinearities are important, such as Bjorken flow. Nevertheless, the results reported here might be useful both in estimating phenomenological
roles and sizes of higher gradient effects as suggested in~\cite{0704.1647,0905.4069} and  in direct studies of experimentally observed phenomena where linear dissipative terms play an important role. Sonic booms created by jets or heavy quarks, fluctuations in the flows and correlations are examples of applications which we have in mind.

A question of convergence of the gradient expansion has been raised in~\cite{1302.0697},
and it has been argued there that radius of convergence is in fact zero presumably due to a factorial growth in the number of terms at high orders. We have not observed any
convergence issues, thus indirectly confirming the conclusions of~\cite{1302.0697} about a nonlinear origin of the convergence problem.

Throughout the paper, we have been referring  to equations~(\ref{abcd eqs}) as holographic RG flow-type equations for the viscosity functions.
Indeed, the radial coordinate $r$ is frequently associated with a scale of RG flow of the boundary CFT. While we do have an evolution in $r$,
we identify physical quantities (viscosities) only at the infinite boundary. In the spirit of holographic Wilsonian RG~\cite{1006.1902,1009.3094,1010.1264,1010.4036,1102.4477,1107.2134}, it would be proper to introduce a finite cutoff along radial direction and define associated physical quantities, such as $\eta(r,\omega,q^2)$ and $\zeta(r,\omega,q^2)$, at the cutoff surface.  That would result in an RG evolution of these momenta-dependent coefficients, thus extending previous results on RG flows of the shear viscosity coefficient $\eta_0$.  So far, the universality of the ratio~(\ref{ratio}) was clarified as a consequence of no RG flow from the horizon to the boundary of the Navier-Stokes hydrodynamics~\cite{0809.3808}.

As further development of this project, we plan to extend our present study to conformal fluids in a weakly curved background manifold, with all order derivative terms resummed. Metric perturbations at the boundary may be taken into account following~\cite{0806.0006,0809.4272}. Additional transport coefficient functions  associated with the boundary curvature are expected to emerge~\cite{0905.4069}.
\acknowledgments
YB would like to thank Yun-Long Zhang for discussions on fluid/gravity correspondence and Jiajun Ma for useful discussion on numerical calculations. ML thanks Edward Shuryak for  early collaborative works that lead to this project. ML is also grateful to the Physics Department of the University of Connecticut for hospitality during the period when this publication was completed. This work was supported by the ISRAELI SCIENCE FOUNDATION grant \#87277111, BSF grant \#012124, the People Program (Marie Curie Actions) of the European Union's Seventh Framework under REA grant agreement \#318921; and the Council for Higher Education of Israel under the PBC Program of Fellowships for Outstanding Post-doctoral Researchers from China and India (2013-2014).

\providecommand{\href}[2]{#2}\begingroup\raggedright\endgroup

\end{document}